\documentclass[preprint,aps ,nofootinbib]{revtex4}

\pdfoutput=1
\usepackage{graphicx}
\usepackage{epsfig}
\usepackage{amsmath}
\usepackage{amsfonts}
\usepackage{amssymb}
\usepackage{color}%
\usepackage{dcolumn}
\usepackage{slashed}
\usepackage{hyperref}
\usepackage{enumerate}
\usepackage{float}

\def\arXiv#1{\href{http://arxiv.org/abs/#1}{arXiv:#1}}
\def\arXiv#1#2{\href{http://arxiv.org/abs/#1}{arXiv:#1}}
\def\arXivid#1#2{\href{http://arxiv.org/abs/#1/#2}{#1/#2}}
\def\doi#1#2{\href{http://doi.org/#1}{#2}}

\providecommand{\U}[1]{\protect\rule{.1in}{.1in}}

\newcommand{\f}{\begin{equation}}
\newcommand{\ff}{\end{equation}}
\newcommand{\fa}{\begin{eqnarray}}
\newcommand{\ffa}{\end{eqnarray}}

\begin{document}
\title{Spectral Weight Suppression and Fermi Arc-like Features \\ with Strong Holographic Lattices}
\author{Sera Cremonini $^{a}$}
\email{cremonini@lehigh.edu}
\author{Li Li $^{b}$}
\email{liliphy@itp.ac.cn}
\author{Jie Ren $^{c}$}
\email{renjie7@mail.sysu.edu.cn}
\affiliation{$^a$ Department of Physics, Lehigh University, Bethlehem, PA, 18018, USA.\\
$^b$CAS Key Laboratory of Theoretical Physics, Institute of Theoretical Physics, Chinese Academy of Sciences, Beijing 100190, China.\\
$^c$School of Physics, Sun Yat-Sen University, Guangzhou, 510275, China.
 }

\begin{abstract}
We investigate holographic fermions in uni-directional striped phases, where the breaking of translational invariance can be generated either spontaneously or explicitly.
We solve the Dirac equation for a probe fermion in the associated background geometry.
When the spatial modulation effect becomes sufficiently strong, we see a 
spectral weight suppression whenever the Fermi surface is larger than the first Brillouin zone.
This leads to the gradual disappearance of the Fermi surface along the symmetry breaking direction, in all of the cases we have examined. 
This effect appears to be a generic consequence of strong inhomogeneities, independently of whether translational invariance is broken spontaneously or explicitly.
The resulting Fermi surface is segmented and has features reminiscent of Fermi arcs.
\end{abstract} \maketitle

\tableofcontents

\newpage

\section{Introduction}

Angle-resolved photoemission spectroscopy (ARPES) and scanning tunneling microscopy (STM) experiments 
are valuable tools for investigating strongly correlated electrons in solids, such as topological materials and unconventional superconductors. 
They trace properties of single-particle spectral functions, 
revealing information about the electronic band structure, the quasi-particle excitation spectrum and the Fermi surface in a material. 
Thus, the comparison of ARPES and STM measurements to spectral functions computed with holographic methods provides a test of holography as a viable framework for describing strongly correlated electron matter. 
In turn, holographic computations have the potential to shed light on the fundamental mechanisms driving such systems.

The earliest studies of fermionic response in holography focused on 
homogeneous cases which respect translational invariance (see~\cite{Henningson:1998cd, Mueck:1998iz, Lee:2008xf, Liu:2009dm, Cubrovic:2009ye, Faulkner:2009wj} for pioneering works and \emph{e.g.}~\cite{Iqbal:2011ae} for a review). 
However, to describe real materials characterized by strong lattice potentials one must work with setups in which translational symmetry is broken. 
Several mechanisms for encoding lattice effects have been put forth within holography,
with the simplest one being the homogeneous holographic lattices of ~\cite{Donos:2012js,Donos:2013eha,Andrade:2013gsa}, which simulate the effects of translational symmetry breaking while retaining the homogeneity of the spacetime geometry.
However, homogeneous lattices are unable to capture 
the physics of Umklapp, motivating the need to work with periodic lattices, as argued by~\cite{Bagrov:2016cnr}.
This was done first in~\cite{Liu:2012tr} through a perturbatively small periodic modulation of the chemical potential 
and later generalized to the fully backreacted case in~\cite{Ling:2013aya}. 
Both of these studies identified an anisotropic Fermi surface and the appearance of a band gap at the Brillouin zone boundary. However, in both of these constructions the lattice periodicity was put in by hand and was irrelevant in the infrared. Nevertheless, a number of striped quantum phases are generated spontaneously in strongly correlated electron systems~\cite{Fradkin2015}, such as charge density wave (CDW), spin density wave (SDW) and pair density wave (PDW) phases. These spontaneous orders are believed to play an important role in the rich phase diagram of strongly correlated systems, and in particular in high temperature superconductors.

Building on this work, we recently extended~\cite{Cremonini:2018xgj} the analysis of fermionic spectral functions 
to the case in which translational invariance is broken spontaneously -- with and without an underlying ionic lattice.
Thus, our construction in~\cite{Cremonini:2018xgj} provided a holographic realization of spontaneous crystallization in the presence of a background lattice, and allowed us to ask under which conditions a Fermi surface forms and how it 
evolves in the presence of strong correlations.
The most intriguing feature that emerged from~\cite{Cremonini:2018xgj} was the suppression of the fermionic spectral function 
once the translational symmetry breaking became sufficiently strong.
In particular, for large enough values of the amplitude of the ionic lattice the Fermi surface gradually disappeared -- along the direction of broken translational symmetry -- leaving behind disconnected segments.
This phenomenon is reminiscent of the \emph{Fermi arcs}~\cite{Norman,Kanigel1,Kanigel2} observed experimentally in the pseudo-gap phase of the high temperature superconductors, suggesting that the Fermi surface in these phases may be open 
and segmented. The spectral function suppression and Fermi arc-like features have also been observed in a homogeneous background by considering a strong Q-lattice deformation in one spatial direction~\cite{Ling:2014bda} or by introducing a dipole-couling between bulk fermionic and gauge fields~\cite{Vanacore:2015poa,Chakrabarti:2019gow}.

In the temperature range examined in~\cite{Cremonini:2018xgj}, the fermionic spectral weight suppression 
was observed in the striped phase only in the presence of an ionic lattice, 
\emph{i.e.} when the breaking of translations is explicit. 
Unfortunately, because of numerical challenges, in the purely spontaneous symmetry breaking case in~\cite{Cremonini:2018xgj} we could not reach temperatures low enough to see the analogous effect.
However, our expectation was that the spectral weight suppression is a general property in holography 
and therefore that it should not be very sensitive to the specific type of spatial modulation present in the system. 
At low temperatures, a sufficiently strong spontaneously generated periodic lattice should also lead to 
such a suppression.

Here we explore this question in further detail by constructing the fermionic spectral function in 
several holographic models of lattices and (striped) phases with broken translational invariance.
Our goal is to ask whether the spectral weight suppression is a generic effect, independently of whether translations are broken spontaneously or explicitly.
To this end, we examine the structure of the Fermi surface and its dependence on the fermionic charge and lattice strength.

In the context of Fermi liquid theory, a Fermi surface can be reconstructed by identifying
singularities in the spectral function $A(\omega,\textbf{k})$ at $\textbf{k}=\textbf{k}_F$ and $\omega=0$ (with the energy defined relative to the chemical potential), at zero temperature. 
The poles denote low-energy excitations slightly outside or slightly inside the Fermi surface -- quasi-particles and quasi-holes, respectively. 
In strongly correlated systems in which Fermi liquid theory breaks down, one can still adopt a working definition of a Fermi 
surface as the surface in momentum space where there are gapless excitations,
resulting in non-analytic behavior of $A(\omega,\textbf{k})$ at $\textbf{k} = \textbf{k}_F$ and $\omega = 0$ 
(see~\emph{e.g.} the discussion in~\cite{Iqbal:2011ae}).  
Moreover, when one is not strictly at zero temperature, the poles are replaced by peaks that must be sufficiently sharp and narrow 
as compared to the temperature scale in the system (generalized Fermi surface criteria at finite temperature were outlined in~\cite{Cosnier-Horeau:2014qya}). 

In our analysis we have adopted this working definition and identified the location and shape of the Fermi surface by inspecting the behavior of $A(\omega,\textbf{k})$ for $\omega \sim 0$ and singling out the appropriate peak structure. 
In all the examples we have studied the fermionic spectral function is suppressed when the translational symmetry breaking is sufficiently strong, 
independently of the particular symmetry breaking mechanisms. This includes cases in which the symmetry breaking is purely spontaneous. Thus, our study confirms and generalizes the results of~\cite{Cremonini:2018xgj}, and 
provides strong numerical evidence to our previous observation: \emph{in strongly correlated systems Fermi surfaces can be suppressed when the inhomogeneity effect is strong enough}.
Moreover, this phenomenon occurs only when the Fermi surface extends beyond the first Brillouin zone, suggesting that the suppression may be the result of an interaction between Fermi surface branches belonging to different zones. 

The structure of the paper is as follows. Section \ref{SetupSection} describes the gravitational models we will work with and the holographic techniques to 
extract the fermionic spectral functions. 
Secion \ref{SectionNumerics} contains our numerical results using the extended zone scheme and Section \ref{SectionDiscussion} is a summary of results and open questions.
Appendix \ref{Appendix} contains the numerical results adopting a ``folded" representation of the spectral function, 
which might be appropriate to describe the band structure in the reduced zone scheme. The preliminary analysis of the energy distribution of the spectral function is presented in Appendix~\ref{app:energy}.

\section{Holographic Method}
\label{SetupSection}
We will work within  the Einstein-Maxwell-scalar framework,
 and consider models that can be captured by examining various cases of 
\begin{eqnarray}\label{generalmodel}
S&=&\frac{1}{2\kappa_N^2}\int d^{4}x \sqrt{-g} \left[\mathcal{R}+\frac{6}{L^2}+\mathcal{L}_{m}\right], \nonumber \\
\mathcal{L}_{m} &=& -\frac{1}{2}\partial_{\mu}\chi \partial^{\mu}\chi-\frac{Z_A(\chi)}{4}F_{\mu\nu}F^{\mu\nu}-\frac{Z_B(\chi)}{4}\tilde{F}_{\mu\nu}\tilde{F}^{\mu\nu}-\frac{Z_{AB}(\chi)}{2}F_{\mu\nu}\tilde{F}^{\mu\nu}\nonumber\\
&&-\mathcal{K}(\chi)(\partial_\mu\theta-q_A A_\mu-q_B B_\mu)^2-V(\chi)\, ,
\label{actions}
\end{eqnarray}
where $F_{\mu\nu}=\partial_\mu A_\nu-\partial_\nu A_\mu$ and $\tilde{F}_{\mu\nu}=\partial_\mu B_\nu-\partial_\nu B_\mu$ are the field strengths of the $U(1)$ vector fields $A_\mu$ and $B_\mu$, respectively.
The scalar field $\chi$ generically couples to both vectors via a St\"{u}ckelberg term, which is introduced to realize 
the spontaneous breaking of the U(1) symmetry~\cite{Franco:2009yz, Aprile:2009ai,Cai:2012es,Kiritsis:2015hoa}.

We are interested in non-linear solutions describing uni-directional striped black branes, which depend on the holographic radial coordinate $z$ and on one of the spatial coordinate, chosen for concreteness to be $x$. 
We will use the DeTurk trick~\cite{Headrick:2009pv} to construct the corresponding background geometry. The following ansatz suffices to accommodate the solutions of interest,
\begin{eqnarray}\label{ansatzbh}
&&ds^2=\frac{r_h^2}{L^2 (1-z^2)^2}\left[-F(z)Q_{tt} \, dt^2+\frac{4 z^2 L^4 Q_{zz}}{r_h^2 F(z)}\, dz^2+Q_{xx}(dx-2 z(1-z^2)^2Q_{xz}dz)^2+Q_{yy}\, dy^2\right], \nonumber \\
&&\chi=(1-z^2)\phi \, , \qquad A_t=\mu\, z^2 \alpha, \qquad B_t=z^2 \beta,
\end{eqnarray}
where the eight functions $\mathcal{Q}=(\phi,\alpha,\beta,Q_{tt},Q_{zz},Q_{xx},Q_{yy}, Q_{xz})$ depend on both $z$ and $x$. 
By taking 
$\phi=\beta=Q_{xz}=0$ and  $\alpha=Q_{tt}=Q_{zz}=Q_{xx}=Q_{yy}=1$
one recovers the standard AdS-RN black brane solution,
\begin{equation}\label{RNadsz}
\begin{split}
&ds^2=\frac{r_h^2}{L^2 (1-z^2)^2}\left[-F(z)dt^2+\frac{4 z^2 L^4}{r_h^2 F(z)}dz^2+(dx^2+dy^2)\right]\,,\\
&F(z)=z^2\left[2-z^2+(1-z^2)^2-\frac{L^2 \mu^2}{4 r_h^2}(1-z^2)^3\right],\quad A_t=\mu\, z^2 ,
\end{split}
\end{equation}
which will be chosen to be the reference metric for the DeTurck method. 
In the present coordinate system the horizon is located at $z=0$ and the AdS boundary at $z=1$.
The black brane temperature is given by
\begin{equation}
\label{temp}
T=\frac{r_h}{4\pi}\left[\frac{3}{L^2}-\frac{\mu^2}{4r_h^2}\right] \,.
\end{equation}

To study the fermionic response, we introduce a probe Dirac fermion $\zeta$ with charge $q$ and a field dependent mass term $M(\chi)$, 
\begin{eqnarray}\label{actionfermion}
S_{D}=i\int d^{4}x \sqrt{-g}\,\overline{\zeta}\left[\Gamma^{\underline{a}}e_{\underline{a}}^\mu\,(\partial_{\mu}+\frac{1}{4}(\omega_{\underline{ab}})_{\mu}\Gamma^{\underline{ab}}-iqA_{\mu})-M(\chi)\right]\zeta \, .
\end{eqnarray}
While throughout the analysis we will consider mostly cases in which the fermion mass is constant, we will also examine a model for which $M(\chi)=m+n \chi^2$ with $m$ and $n$ constants.
Here $(\mu,\nu)$ and $(\underline{a},\underline{b})$ denote, respectively, the bulk spacetime and tangent indices, $\Gamma^{\underline{a}}$ are gamma matrices with $\Gamma^{\underline{bc}}=\frac{1}{2}[\Gamma^{\underline{b}},\Gamma^{\underline{c}}]$ and $e_{\underline{a}}^\mu$ is the vielbein, with $(\omega_{\underline{ab}})_{\mu}=e_{\underline{a}\nu}\nabla_\mu e_{\underline{b}}^\nu$ the associated spin connection.
We can obtain the retarded Green's function for the fermionic operator of the strongly coupled field theory by solving the bulk Dirac equation with in-falling boundary conditions at the horizon. 

For the phase described by the striped geometry~\eqref{ansatzbh}, following the same choice of vielbein and gamma matrixes of~\cite{Cremonini:2018xgj} we obtain the Dirac equation 
\begin{eqnarray} \label{Dirac1}
\left(\partial_{z}+2z(1-z^2)^2Q_{xz}\partial_x+\Pi_{1}\mp\frac{2 M L z}{1-z^2}\sqrt{\frac{Q_{zz}}{F}}\right)
\left[ \begin{matrix} \mathcal{A}_{1} \cr  \mathcal{B}_{1} \end{matrix}\right]
&\pm& \Pi_{2}\left[ \begin{matrix} \mathcal{B}_{1} \cr  \mathcal{A}_{1} \end{matrix}\right]\\\nonumber
+ i\frac{2L^2z}{r_h}\sqrt{\frac{Q_{zz}}{F Q_{xx}}}(\partial_x+\Pi_3)\left[ \begin{matrix} \mathcal{B}_{1} \cr  \mathcal{A}_{1} \end{matrix}\right]
&+&\frac{2L^2z}{r_h}\sqrt{\frac{Q_{zz}}{F Q_{yy}}}k_y \left[ \begin{matrix} \mathcal{B}_{2} \cr  \mathcal{A}_{2} \end{matrix}\right]
=0,
\end{eqnarray}
\begin{eqnarray} \label{Dirac2}
\left(\partial_{z}+2z(1-z^2)^2Q_{xz}\partial_x+\Pi_{1}\mp\frac{2 M L z}{1-z^2}\sqrt{\frac{Q_{zz}}{F}}\right)
\left[ \begin{matrix} \mathcal{A}_{2} \cr  \mathcal{B}_{2} \end{matrix}\right]
&\pm&\Pi_{2}\left[ \begin{matrix} \mathcal{B}_{2} \cr  \mathcal{A}_{2} \end{matrix}\right]\\\nonumber
- i\frac{2L^2z}{r_h}\sqrt{\frac{Q_{zz}}{F Q_{xx}}}(\partial_x+\Pi_3)\left[ \begin{matrix} \mathcal{B}_{2} \cr  \mathcal{A}_{2} \end{matrix}\right]
&+&\frac{2L^2z}{r_h}\sqrt{\frac{Q_{zz}}{F Q_{yy}}}k_y \left[ \begin{matrix} \mathcal{B}_{1} \cr  \mathcal{A}_{1} \end{matrix}\right]
=0,
\end{eqnarray}
with
\begin{equation}
\begin{split}
\Pi_1&=z(1-z^2)^2[2 i \,k_x\, Q_{xz}(z,x)+\partial_x Q_{xz}(z,x)]\,,\\
\Pi_2&=\frac{2L^2z}{r_h F(z)}\sqrt{\frac{Q_{zz}(z,x)}{Q_{tt}(z,x)}}[\omega+q\mu z^2\,\alpha(z,x)]\,,\\
\Pi_3&=i\, k_x-\frac{\partial_x Q_{xx}(z,x)}{4 Q_{xx}(z,x)}+\frac{\partial_x Q_{zz}(z,x)}{4 Q_{zz}(z,x)}\,.
\end{split}
\end{equation}
Note that we have made the transformation
\begin{equation}
\zeta=\left(\frac{L(1-z^2)}{r_h}\right)^{3/2}(FQ_{tt}Q_{xx}Q_{yy})^{-1/4}\left[\begin{matrix} \Psi_{1} \cr  \Psi_{2} \end{matrix}\right]\,,
\end{equation}
and have expanded each two-component spinor, 
\begin{eqnarray}
\Psi_{\alpha}(t,z,x,y) \sim \left[ \begin{matrix} \mathcal{A}_{\alpha}(z,x) \cr  \mathcal{B}_{\alpha}(z,x) \end{matrix}\right] e^{-i\omega t+k_x x+k_y y},\quad \alpha=1, 2\,.
\end{eqnarray}

Taking into account the periodicity of the background 
along the $x$ direction,  which is fixed by an ``Umklapp wave vector" $K$, the solution to the Dirac equation can be expanded using Bloch's theorem as 
\begin{equation}
\mathcal{A}_{\alpha}(z,x)=\sum_{n=0,\pm1,\pm2,\cdots} \mathcal{A}_{\alpha}^{(n)}(z)\, e^{i n K x},\qquad \mathcal{B}_{\alpha}(z,x)=\sum_{n=0,\pm1,\pm2,\cdots} \mathcal{B}_{\alpha}^{(n)}(z)\, e^{i n K x},
\end{equation}
with $n$ characterizing the momentum level or Brillouin zone.
Near the $z=1$ AdS boundary the asymptotic expansion is of the form 
\begin{eqnarray} \label{boundaryn}
\left[\begin{matrix} \mathcal{A}_{\alpha}^{(n)} \cr  \mathcal{B}_{\alpha}^{(n)} \end{matrix}\right]
 {=} a_{\alpha}^{(n)}(1-z)^{-mL}\left[ \begin{matrix} 1 \cr  0 \end{matrix}\right]
+b_{\alpha}^{(n)}(1-z)^{mL}\left[\begin{matrix} 0 \cr 1 \end{matrix}\right]+\cdots\,,
\end{eqnarray}
where $(a_{\alpha}^{(n)}, b_{\alpha}^{(n)})$ are constants for a given $(\omega, k_x, k_y)$ and we have assumed that the mass term $M(\chi)$ approaches a constant $m$ at the AdS boundary.
The retarded Green's function can then be obtained from the following relation~\cite{Liu:2012tr,Ling:2013aya},
\begin{equation}
b_{\alpha}^{(n)}(\omega, k_x, k_y)=\sum_{\alpha',n'}G^R_{\alpha,n;\alpha',n'}(\omega, k_x, k_y)\,a_{\alpha'}^{(n')}(\omega, k_x, k_y)\, ,
\end{equation}
from which one extracts~\cite{Cremonini:2018xgj} the spectral density (\emph{a.k.a.} the spectral function)  $A(\omega, k_x,k_y)$.

To provide a representation of the band structure in the extended zone scheme 
we work with the following ``unfolded'' expression for the spectral function,
\begin{equation}\label{newspectral}
A(\omega, k_x=k_0+n K, k_y)=\text{Tr}\,\text{Im}[G^R_{\alpha,n;\alpha',n}(\omega, k_0,k_y)]\,,
\end{equation}
with $k_0\in[-\frac{K}{2},\frac{K}{2}]$ restricted to the first Brillouin zone, but $k_x$ ranging over the entire system.
The integer $n$ denotes once again the momentum level or Brillouin zone number. 
Since this spectral function representation doesn't implement any folding procedure, it will allow for a more direct 
comparison of our results with the ARPES measurements, which are a direct probe of the
electronic band structure in the extended Brillouin zone scheme. 
We will use (\ref{newspectral}) in the main body of the paper, to facilitate the discussion of the features which are reminiscent 
of Fermi arcs.
On the other hand, for completeness in Appendix A we will adopt a different expression for the spectral function, 
which provides a description of the band structure in the reduced zone scheme and is given by
\begin{equation}\label{spectral}
A(\omega, k_x,k_y)=\sum_{n=0,\pm1,\pm2,\cdots}\text{Tr}\; \text{Im}[G^R_{\alpha,n;\alpha',n}(\omega, k_x,k_y)]\,,
\end{equation}
where $k_x\in[-\frac{K}{2},\frac{K}{2}]$ is restricted to be in the first Brillouin zone.
By summing over all values of $n$, we ensure that all the branches outside the first Brillouin zone are folded back into it. 
As we will show in this paper, our main result, \emph{i.e.} the spectral weight suppression in the presence of 
strong lattice inhomogeneity, is independent of which representation we adopt.

\section{Numerical Results}
\label{SectionNumerics}

We are now ready to compute the spectral density (\ref{newspectral}) in the presence of strong spatial modulations, for various holographic models described by different cases of (\ref{generalmodel}).  
We refer the reader to~\cite{Cremonini:2018xgj} for further details on the numerical procedure used in this analysis. 
In the present study we will work in the grand canonical ensemble with the chemical potential $\mu=1$ and fix the AdS radius to be $L=1/2$. 
We will take $m=0$ and consider different values of the charge $q$ as well as the mass term $n$ for the spontaneous symmetry breaking case, which corresponds to scanning through different dual boundary field theories. 

\subsection{Explicit Lattice}
\label{explicit_case}

We start by examining the case in which translational invariance is broken explicitly by turning on a source in the ultraviolet (UV) of the theory. First, we consider pure Einstein-Maxwell theory with a spatially modulated chemical potential which mimics  
a harmonic potential, thus playing the role of an ionic lattice.
We then move on to a simple Einstein-Maxwell-scalar model with a spatially inhomogeneous 
boundary condition for the scalar field, describing a scalar lattice.
In both instances we will see a sharp decline in the fermionic spectral density when the lattice amplitude is increased sufficiently.

\subsubsection{Ionic lattice}
\label{ioniclatticesubsection}

The simplest setup we can consider is the standard Einstein-Maxwell theory 
\begin{equation}
\mathcal{L}_m=-\frac{1}{4}F_{\mu\nu}F^{\mu\nu}\,,
\end{equation}
which is obtained from the large class of models (\ref{generalmodel}) by consistently setting $\chi=B_\mu=0$ with $Z_A=1$ and $Z_B=Z_{AB}=\mathcal{K}=V=0$. 
The ionic lattice is introduced through a spatially varying chemical potential, which can be viewed as representing the potential felt by electrons in an array of ions.
In holography the chemical potential is identified with the boundary value of the temporal component $A_t$ of the gauge field. At the AdS boundary $z=1$, we choose
\begin{equation}\label{ioniclattice}
\mu(x)=A_t(z=1,x)=\mu[1+a_0 \cos(p_I\, x)]\,,
\end{equation}
which describes a uni-directional single harmonic potential with wave vector $p_I$ and relative amplitude $a_0$. 

\begin{figure}[ht!]
\begin{center}
\includegraphics[width=.49\textwidth]{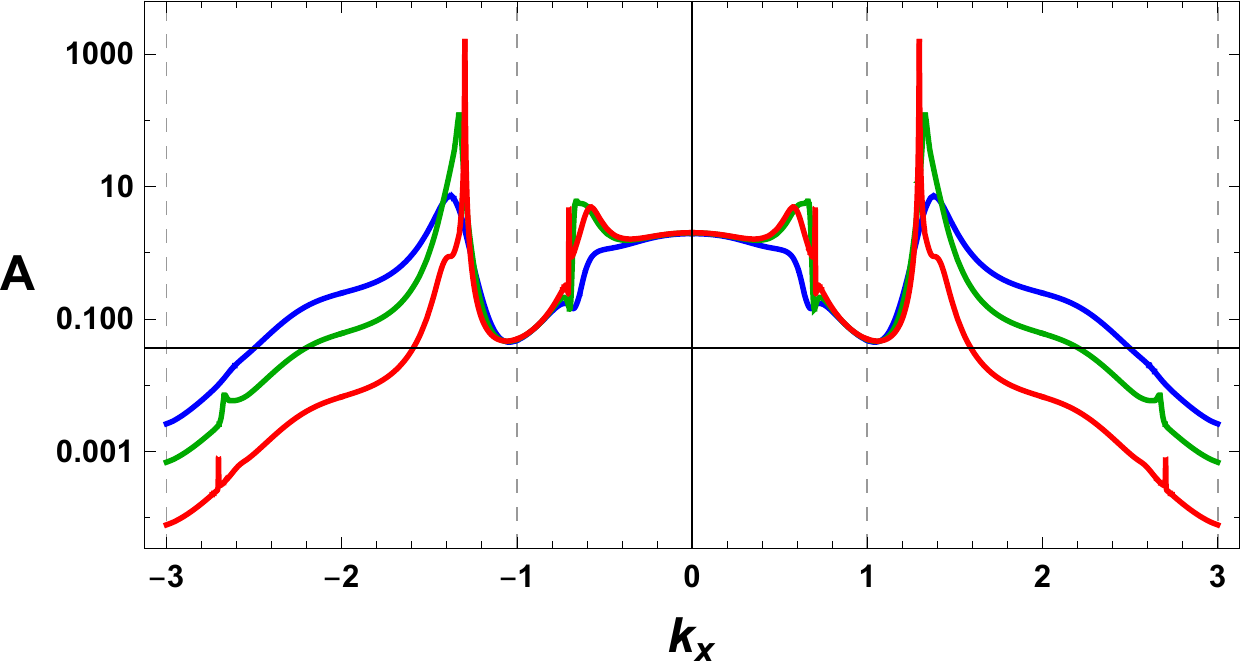}\quad
\includegraphics[width=.48\textwidth]{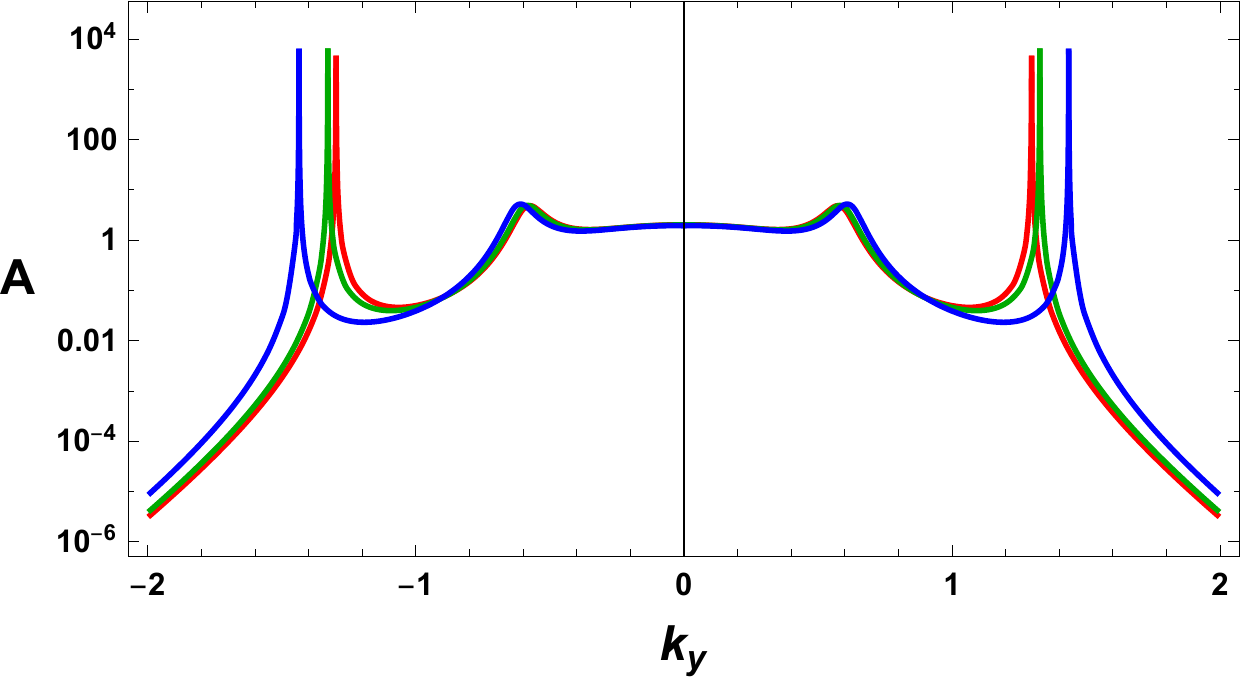}
\caption{Momentum distribution of the spectral density along the $k_x$ axis (left panel) and the $k_y$ axis (right panel) for varying values of the amplitude of the ionic lattice. 
Translational invariance is broken explicitly in the $x$ direction and the first Brillouin zone boundary is at $k_x=\pm1$. 
The red, green and blue curves correspond to $a_0=0.1, 0.3$ and $0.6$, respectively. 
We have fixed $\omega=10^{-6}$, $q=2$, $T=0.0069$ and wave vector $p_I=2$. 
Since the vertical axis is logarithmic, the red curve describes a very sharp peak which is indicative of a Fermi surface.}
\label{fig:Avskionicq20}
\end{center}
\end{figure}

\begin{figure}[ht!]
\begin{center}
\includegraphics[width=.49\textwidth]{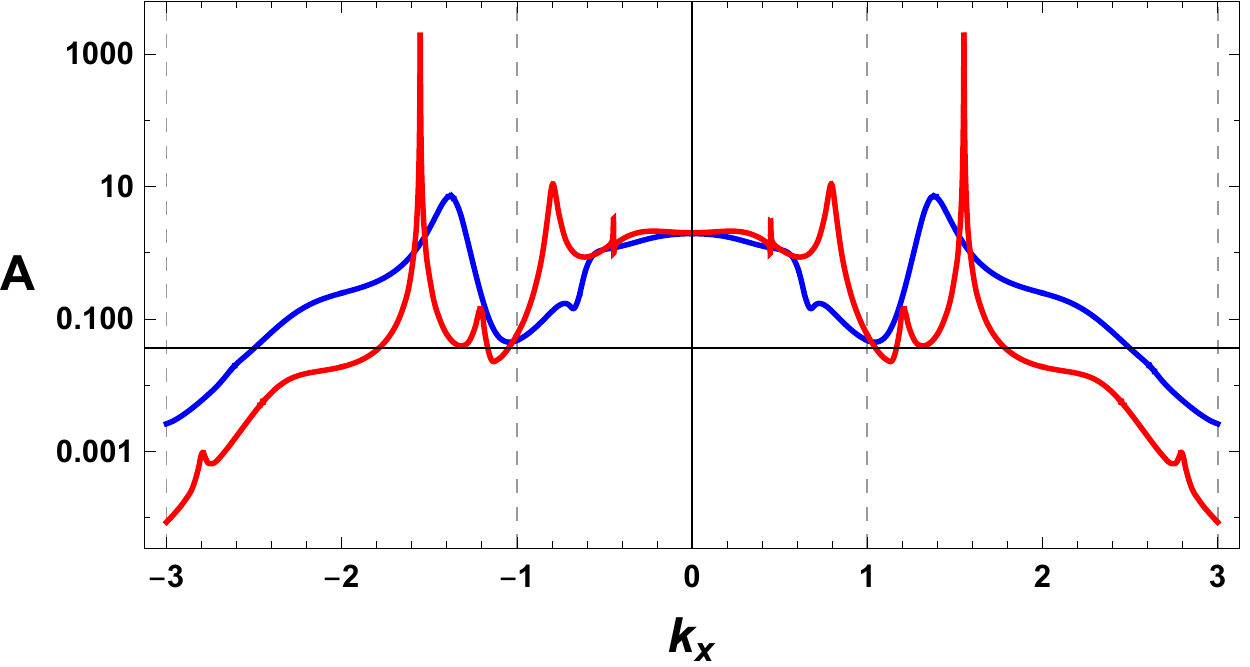}\quad
\includegraphics[width=.48\textwidth]{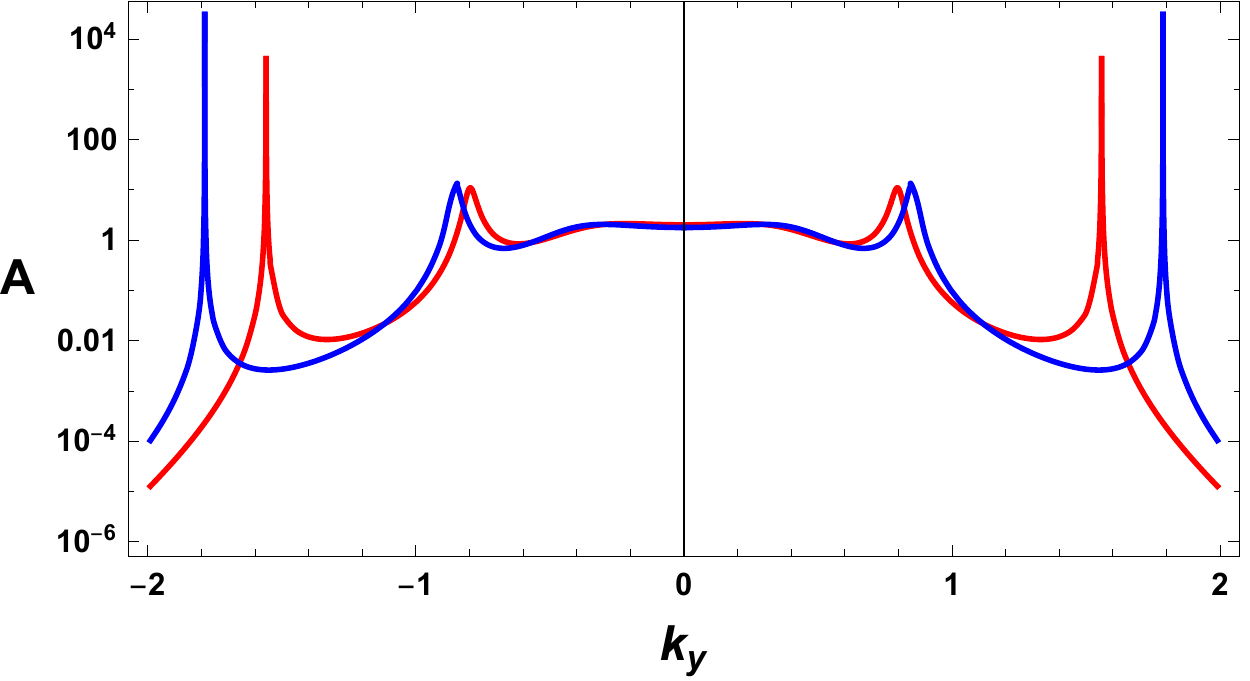}
\caption{Momentum distribution of the spectral density for $q=2.3$ 
for lattice amplitudes $a_0=0.1$ (red curve) and $a_0=0.6$ (blue curve).  The vertical axis is logarithmic. 
The remaining parameters are chosen as in Fig.\,\ref{fig:Avskionicq20}. 
}
\label{fig:Avskionicq23}
\end{center}
\end{figure}

The momentum distribution of the spectral density is shown in Fig.\,\ref{fig:Avskionicq20} for $q=2$ and Fig.\,\ref{fig:Avskionicq23} for $q=2.3$, near zero frequency, $\omega = 10^{-6}$. 
One immediately finds that the spectral density~\eqref{newspectral} is \emph{not} periodic in the extended zone. 
However, there is a particularly interesting feature visible from our numerics: when the peaks appearing in each Brillouin zone in the extended zone representation~\eqref{newspectral} are sufficiently sharp,  they differ from each other by the Umklapp wave vector $K$. 
This ceases to be the case when the lattice effects become sufficiently strong.
As a typical example, let's consider the ionic lattice case with $a_0=1$ (red curve) in the left plot of Fig.\,\ref{fig:Avskionicq20}. The sharpest, dominant peak (which we would associate with a point on the Fermi surface) is located at $k_x\approx -1.297$ ($k_x\approx1.297$). There are additional peaks that are sharp but have a much smaller amplitude\,\footnote{It would be interesting to track their behavior at $T=0$, \emph{i.e.} in the ground state geometry, to determine if they persist and become sharper, or are no longer present.}, located at $k_x\approx0.703$ ($k_x\approx-0.703$) and $k_x\approx2.703$ ($k_x\approx-2.703$). 
Interestingly, they differ from the Fermi momentum $k_x\approx-1.297$ ($k_x\approx1.297$) by the Umklapp wave vector $K=2$.
Next, consider the green curve corresponding to a larger amplitude, $a_0=0.3$. 
We now see smoother peaks located at $k_x\approx-1.329, k_x\approx0.661$ and $k_x\approx 2.667$, and the distance between them is slightly different from the Umklapp wave vector $K=2$. 
For the even larger lattice $a_0=0.6$ (the blue curve in Fig.\,\ref{fig:Avskionicq20}), one can only see two smooth bumps centered at $k_x\approx\pm 1.382$.
Thus, when the lattice effect is large enough
we lose the feature of secondary peaks differing from the Fermi 
momentum (the dominant peak) by an integral multiple of the Umklapp wave vector $K$.

The fact that the fermionic spectral function in the extended zone is not necessarily periodic was already noted in~\cite{Liu:2012tr}, which argued that such non-periodicity 
is expected for non-Fermi liquids. Indeed, for the latter single particle momentum is not
a meaningful quantum number and thus the system need not be
precisely periodic\,\footnote{We thank Jan Zaanen for pointing this out to us.}  in the extended single particle momentum 
Brillouin zone (unlike for free fermions and Fermi liquids, for which the periodicity of
the potential in real space leads to a perfect periodicity in momentum space in the
structure of the extended Brillouin zone).
Thus, the non-periodicity is a generic feature and a direct probe of the 
non-Fermi liquid nature of electron matter, and it would be very interesting to test it experimentally.

Let's now discuss the main new feature of our work, the spectral density suppression and Fermi surface segmentation.
In Fig.\,\ref{fig:Avskionicq20} we see a sharp peak characteristic of a Fermi surface when the amplitude of the ionic lattice is fairly small ($a_0=0.1$). As the lattice amplitude increases, we find a sharp reduction of spectral weight along the $x$-axis, the direction of broken translational symmetry, while the peaks along the $y$-direction remain roughly unchanged.
When the fermionic charge is increased sufficiently, additional peaks start to develop, which would eventually turn into new Fermi surfaces if the charge became large enough. 
This behavior is more visible in Fig.\,\ref{fig:Avskionicq23}, where a secondary peak appears, too small and smooth to indicate a formed Fermi surface.  As in Fig.\,\ref{fig:Avskionicq20}, here we also see a clear disappearance of the spectral weight with increasing
lattice strength. From the behavior of the two peaks in Fig.\,\ref{fig:Avskionicq23}, the Fermi surface suppression appears to be unrelated to the presence of the (growing) secondary peak (see also the discussion in~\cite{Cremonini:2018xgj}).

\begin{figure}[ht!]
\begin{center}
\includegraphics[width=.485\textwidth]{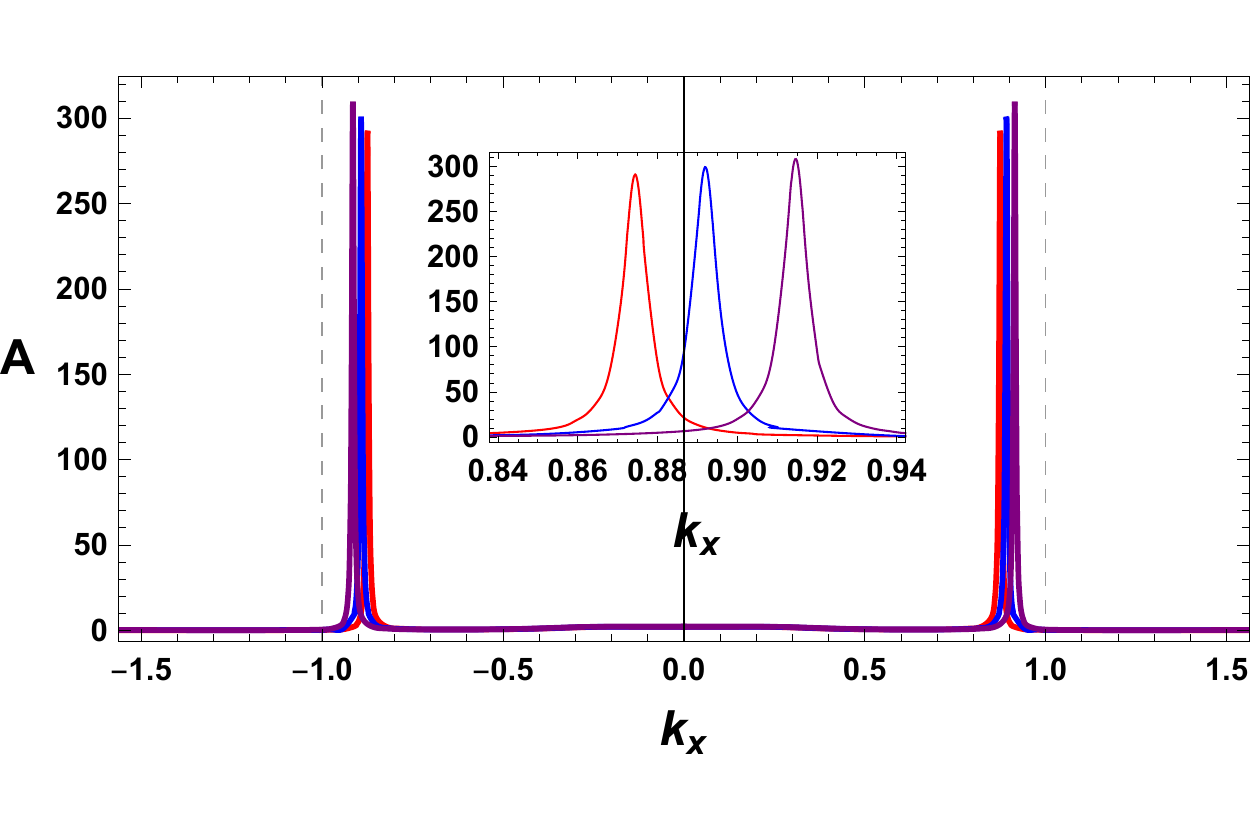}\quad
\includegraphics[width=.485\textwidth]{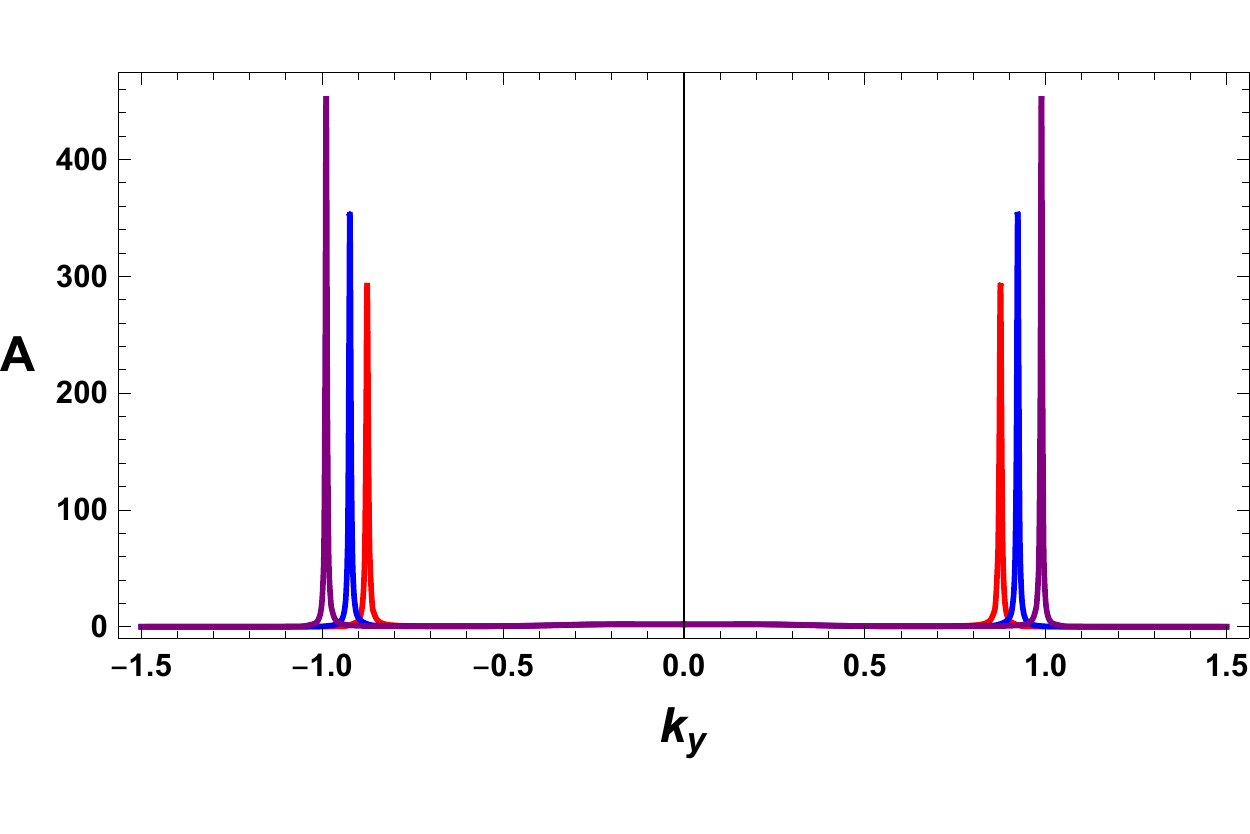}
\caption{Momentum distribution of the spectral density for $q=1.5$ 
for ionic lattice amplitudes $a_0=0.1$ (red curve), $a_0=0.6$ (blue curve) and $a_0=0.9$ (purple curve).  
All remaining parameters were chosen as in Fig.\,\ref{fig:Avskionicq20}. 
The inset in the left panel zooms into the location of the three peaks, which all lie within the 
first Brillouin zone. We see no spectral weight suppression as 
the lattice amplitude is increased.}
\label{fig:Avskionicq15}
\end{center}
\end{figure}

\begin{figure}[ht!]
\begin{center}
\includegraphics[width=.485\textwidth]{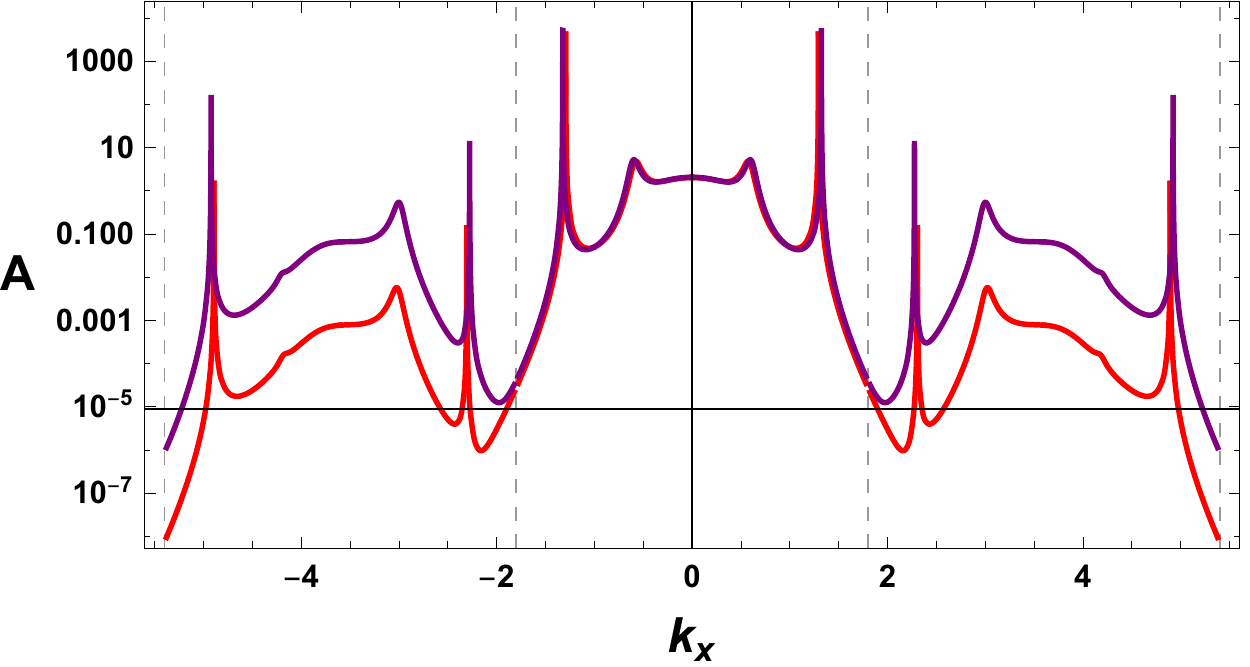}\quad
\includegraphics[width=.485\textwidth]{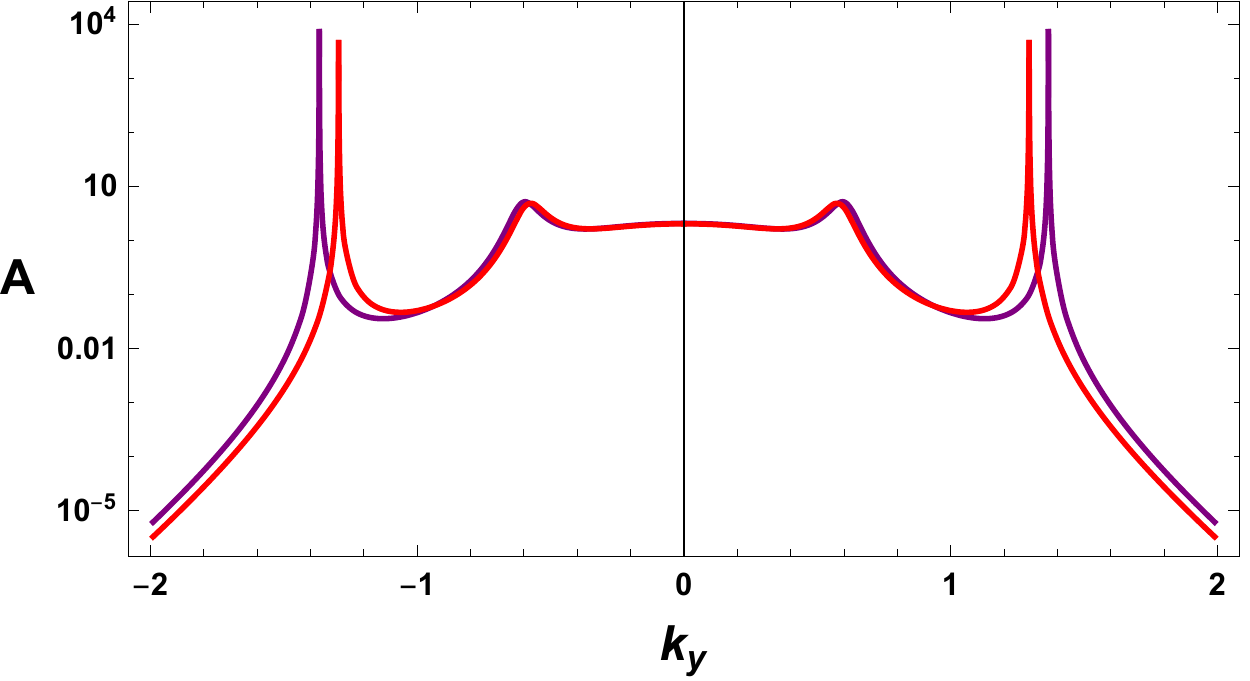}
\caption{Momentum distribution of the spectral density along the $k_x$ axis (left panel) and the $k_y$ axis (right panel) for varying values of the amplitude of the ionic lattice with the wave vector $p_I=3.6$. 
Translational invariance is broken explicitly in the $x$ direction and the first Brillouin zone boundary is at $k_x=\pm1.8$. 
The red and purple curves correspond to $a_0=0.1$ and $a_0=0.9$, respectively. 
We have fixed $\omega=10^{-6}$, $q=2$, $T=0.0069$. 
Since the vertical axis is logarithmic, the two curves describe very sharp peaks indicative of a Fermi surface.}
\label{fig:Avskionicq20k36}
\end{center}
\end{figure}

In the two cases discussed above the fermionic charge $q$ is large enough so that the Fermi surface crosses the first Brillouin zone.  
One could ask if the spectral weight suppression comes from the interaction between Fermi surface branches belonging to different Brillouin zones. 
To test whether this is the case, we can check if we still find a suppression when 
the Fermi surface is contained entirely within the first Brillouin zone and does not cross it. 
The case corresponding to $q=1.5$, for which the peaks are inside the Brillouin zone\,\footnote{Even though the amplitude of the peaks of Fig.\,\ref{fig:Avskionicq15} is a few times smaller than in the two cases above, it is still sufficient to identify the location of the Fermi surface.}, is shown in Fig.\,\ref{fig:Avskionicq15}. 
It is clear that there is no spectral weight suppression as the lattice amplitude is increased.
An easy way to obtain a sharp spectral density peak that lies completely inside the first Brillouin zone is to broaden the zone itself, 
which can be done by increasing the wave vector $p_I$ of the ionic lattice~\eqref{ioniclattice}. 
When the Brillouin zone is sufficiently large as compared to the Fermi momentum one expects a Fermi surface that is almost isotropic. 
This is indeed the case shown in Fig.\,\ref{fig:Avskionicq20k36}, where we plot the spectral density for $p_I=3.6$ but with the same charge $q=2$ as in Fig.\,\ref{fig:Avskionicq20}.


Comparing the two red curves in Fig.\,\ref{fig:Avskionicq20} and Fig.\,\ref{fig:Avskionicq20k36}, we see that the location of the Fermi surface along the $k_y$ direction is almost unchanged (the Fermi momentum $k_{y}^F\approx1.297$ for $p_I=2$ and $k_{y}^F\approx1.294$ for $p_I=3.6$). This is expected since there is no lattice in the $y$ direction and we have also chosen the same charge and temperature. 
As the amplitude of the ionic lattice is increased up to $a_0=0.9$, for which the Fermi surface is still within the first Brillouin zone, we don't see 
any spectral weight suppression (on the other hand the spectrum is slightly enhanced). 
This is similar to the behavior observed in Fig.\,\ref{fig:Avskionicq15}. 

Although we still do not understand the fundamental physical reasons behind the generic suppression of the spectral weight at large lattice deformation, our  analysis suggests that it might due to some kind of interaction between bands. We shall return to this point later.

\subsubsection{Scalar lattice}
Another example of an explicit lattice can be constructed by restoring the scalar field $\chi$ and 
considering the following matter sector,
\begin{equation}
\mathcal{L}_{m} = -\frac{1}{2}\partial_{\mu}\chi \partial^{\mu}\chi-\frac{1}{4}F_{\mu\nu}F^{\mu\nu}-V(\chi)\,,
\label{scalarlagmain}
\end{equation}
which can be obtained from~\eqref{generalmodel} by turning off $B_\mu$ and choosing the scalar couplings to be $Z_A=1$ and $Z_B=Z_{AB}=\mathcal{K}=0$ .
For numerical convenience, we take the scalar potential to describe a mass term of the form $V=-\frac{2}{L^2}\chi^2$. 
With this choice, the asymptotic behavior of the rescaled scalar $\phi=\chi/(1-z^2)$ introduced in~\eqref{ansatzbh} 
near the AdS boundary is 
\begin{equation}
\phi=\phi_s(x)+(1-z^2)\phi_v(x)+\cdots\, .
\end{equation}
A scalar lattice is then introduced by adopting a spatially inhomogeneous boundary condition for $\phi$. We focus on the single harmonic case by specifying
\begin{equation}\label{scalarlattice}
\phi_s(x)=A_0\cos(p_S\, x)\,,
\end{equation}
which corresponds to a periodic source for the dimension-two scalar operator dual to $\chi$.
\begin{figure}[ht!]
\begin{center}
\includegraphics[width=.485\textwidth]{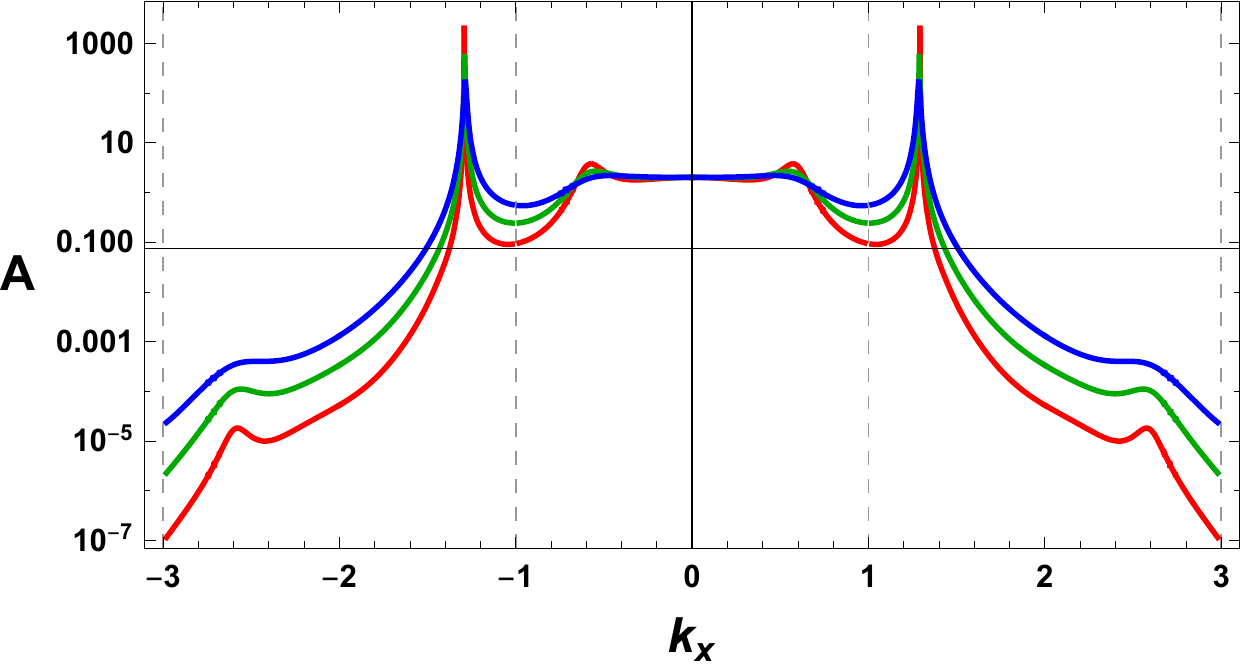}\quad
\includegraphics[width=.485\textwidth]{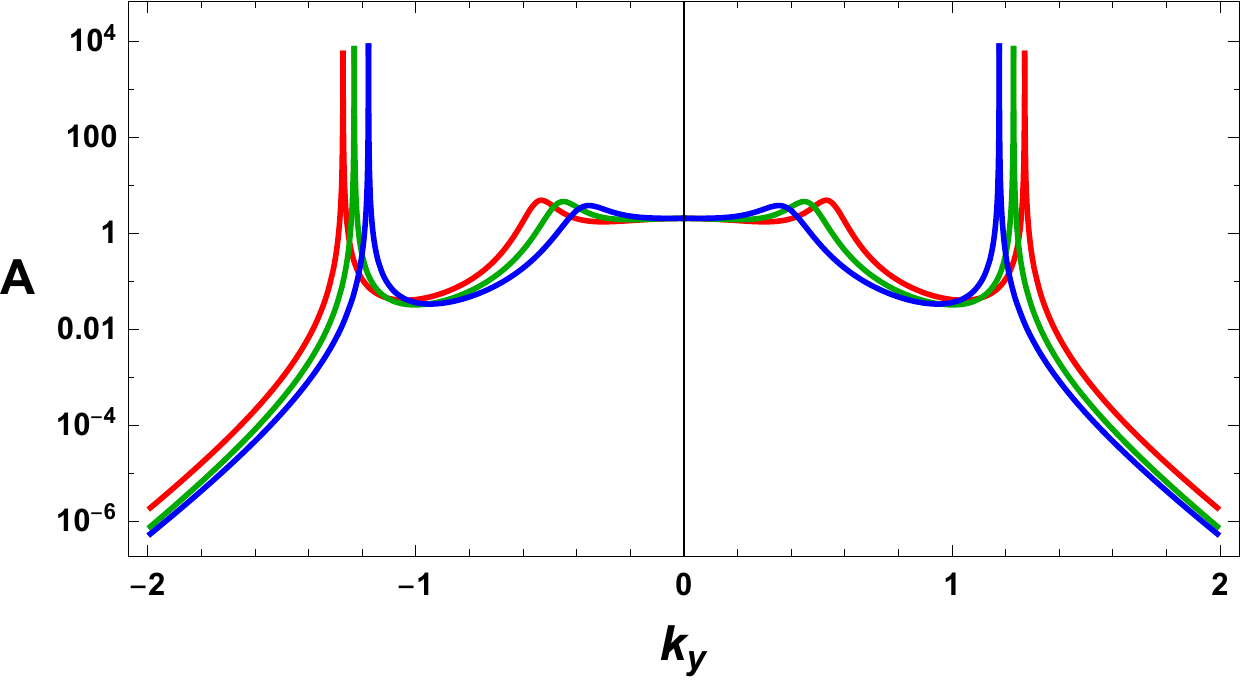}
\caption{Momentum distribution of the spectral density along the $k_x$ axis (left panel) and the $k_y$ axis (right panel) for varying values of the amplitude of the scalar lattice. 
Translational invariance is broken explicitly in the $x$ direction and the first Brillouin zone boundary is at $k_x=\pm1$. 
The vertical axis is logarithmic. 
In both plots the red, green and blue curves correspond to $A_0=3, 6$ and $11$, respectively. 
We have chosen $\omega=10^{-6}$, $q=2$, $T=0.0069$ and wave vector $p_S=1$.}
\label{fig:Avskscalar}
\end{center}
\end{figure}

Fig.\,\ref{fig:Avskscalar} displays the same spectral weight suppression seen in the case of the ionic lattice.
Indeed, the sharp Fermi surface peak visible in the red curve in the $k_x$ momentum distribution broadens and weakens as the lattice strength $A_0$ is increased, until one would no longer interpret the peak as describing a Fermi surface. Once again the structure along the $k_y$ direction remains unchanged, and overall the behavior of the spectral function is analogous to that of the ionic lattice. 
Thus, the analysis of this section indicates that -- independently of which type of explicit lattice one considers --
the spectral function along the symmetry broken direction is suppressed once the amplitude of the lattice is sufficiently large. 
As we observed in our previous paper~\cite{Cremonini:2018xgj}, this suppression then leads to a disconnected Fermi surface and to the appearance of features partially reminiscent of Fermi arcs.

\subsection{Spontaneously Generated Lattice}

Next, we turn our attention to models in which spatial modulations are generated by the spontaneous breaking of translational symmetry, without the use of any explicit source. 
Within the framework of~\eqref{generalmodel} two cases are particularly interesting.
The first one is the realization of a particular example of intertwined orders, 
in which a U(1) symmetry and translational invariance are broken spontaneously at the same time~\cite{Cremonini:2016rbd,Cremonini:2017usb}.
Since in the broken symmetry phase this construction mimics certain features of PDW phases, we will refer to it as the PDW case.\,\footnote{Another holographic PDW model which describes a superconducting phase intertwined with charge, current, and parity orders can be found in~\cite{Cai:2017qdz}.}
The second case describes the spontaneous breaking of translations while retaining the U(1) symmetry, and corresponds to a pure CDW state. 
It can be obtained from~\eqref{generalmodel} by setting $q_A=q_B=0$ and consistently truncating $\theta$~\cite{Donos:2013gda,Ling:2014saa}.

In our previous work~\cite{Cremonini:2018xgj} 
we found -- in the temperature range that could be reached in that analysis -- that the Fermi surface did not display spectral weight suppression when only spontaneous order was considered (\emph{i.e.}, in the ``pure PDW" case). 
In contrast, when adding to the pure PDW construction the ionic UV lattice~\eqref{ioniclattice}, the spectral function along the symmetry broken direction was suppressed once the amplitude of the lattice 
became sufficiently large. This suppression then led to a disconnected Fermi surface~\cite{Cremonini:2018xgj}, as in the cases analyzed in this paper. 
We believe that the difference between the two cases is due to the fact that at the temperatures examined in~\cite{Cremonini:2018xgj} 
the spontaneously generated IR modulation was not strong enough -- as compared to the explicit UV  modulation -- and thus didn't leave an observable imprint. 
Indeed, we expect that at very low temperatures a sufficiently strong spontaneous modulation  would lead to a similar suppression of the fermionic spectral density. 

Constructing the striped background geometries numerically at temperatures low enough to see this effect is particularly challenging, and we haven't been able to do so.
To circumvent this difficulty, we will adopt another mechanism to amplify the effect of the spontaneously generated modulations on the fermion. 
A particularly simple way is to introduce a coupling between the scalar $\chi$ and the fermion $\zeta$. 
Here we consider the coupling $n\chi^2\overline{\zeta}\zeta$, with $n$ a constant, which means 
that the effective mass of the scalar appearing in  (\ref{actionfermion}) is of the form 
\begin{equation}
M(\chi)=m+n\chi^2\,.
\end{equation}
The effect of the spontaneously broken translational symmetry felt by the bulk fermion then becomes stronger as the size $n$ of the interaction term is increased. 
As we will see below, this is sufficient to see the spectral weight suppression.

\subsubsection{Pair Density Wave Case}

We start with a setup that allows us to break spontaneously both the U(1) symmetry and translational invariance -- through the same mechanism and at the same time.
In this case, which we refer to as the PDW, we take the couplings in~\eqref{generalmodel} to be given by
\begin{equation}
\label{PDWcouplings}
\begin{split}
& Z_A(\chi)=1+2\chi^2,\quad Z_B(\chi)=1,\quad Z_{AB}(\chi)=-2.34\, \chi\,, \\
& V( \chi)=-\frac{1}{L^2}\chi^2,\quad \mathcal{K}(\chi)=\frac{1}{2}\chi^2\,,
\end{split}
\end{equation}
with $q_A=1, q_B=0$. As we showed in~\cite{Cremonini:2016rbd,Cremonini:2017usb}, striped order will develop spontaneously below a 
certain critical temperature, with an intrinsic wavelength $k$. As a typical example, we focus on the $k$ =1 branch, with the corresponding critical temperature being $T_c= 0.016$.

%
\begin{figure}[ht!]
\begin{center}
\includegraphics[width=.49\textwidth]{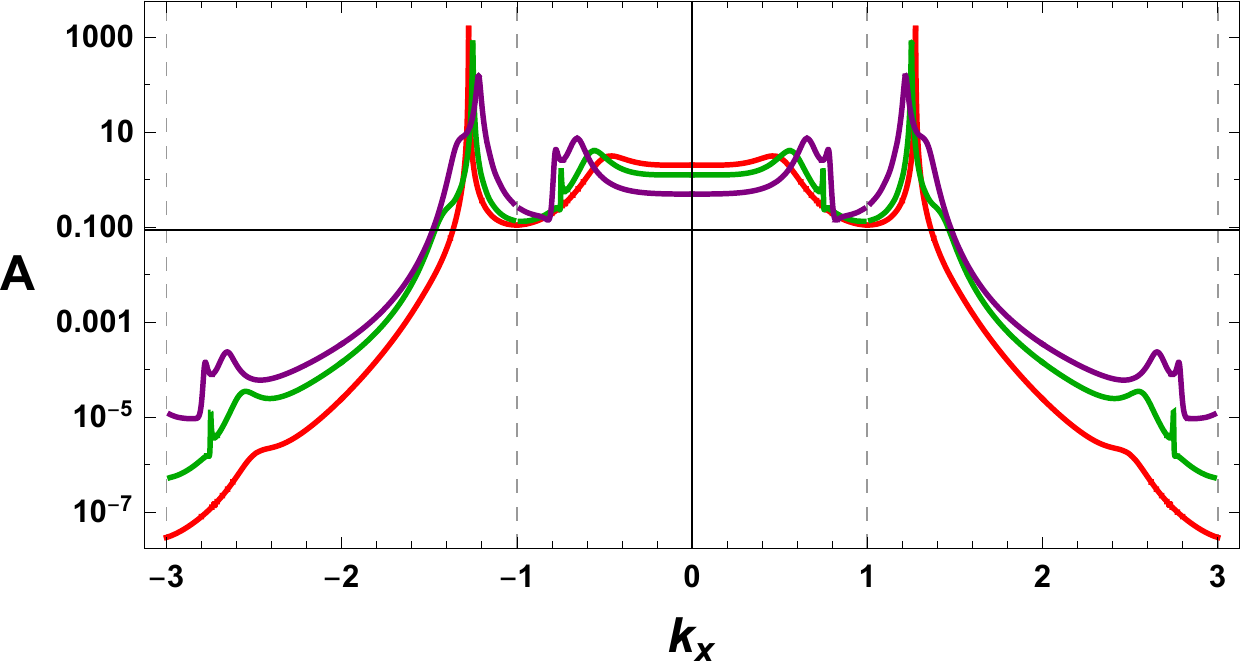}\quad
\includegraphics[width=.48\textwidth]{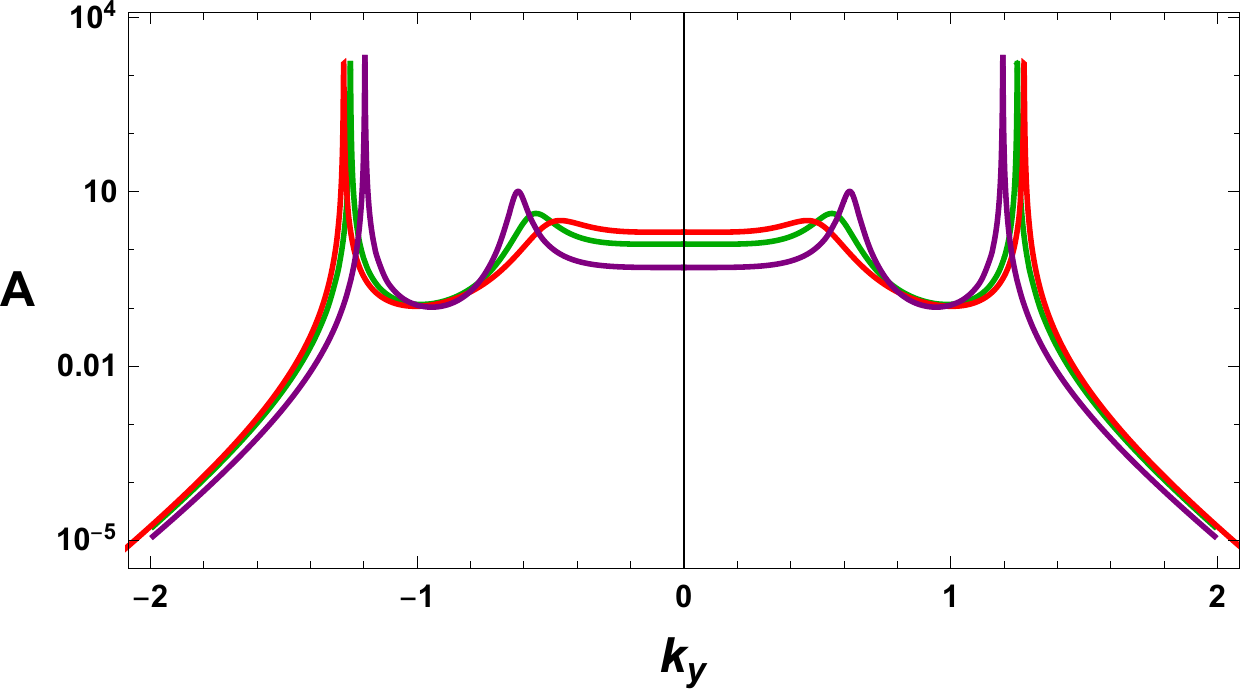}
\caption{Momentum distribution of the spectral density along the $k_x$ axis (left panel) and along the $k_y$ axis (right panel) for varying values of the coupling constant $n$. 
Translational invariance is broken spontaneously along the $x$ direction and the first Brillouin zone boundary is at $k_x=1$. In both plots the red curve corresponds to $n=0$, while the green, and purple curves correspond to $n=10$ and $30$, respectively. We have fixed $\omega=10^{-6}$, $q=2$ and chosen $T=0.014$ for the PDW geometry. }
\label{fig:AvskPDW}
\end{center}
\end{figure}

The corresponding momentum distribution of the spectral density along the symmetry breaking direction is shown in the left plot of Fig.\,\ref{fig:AvskPDW}, for different values of the coupling $n$. 
Very sharp peaks indicative of a Fermi surface are present for small values of $n$. 
However, as the strength of the coupling is increased -- \emph{i.e.} the effect of the spontaneously generated modulation on the fermion becomes large -- the peak of the spectral density becomes weaker and broader. In contrast to the suppression of the spectral density along the symmetry breaking direction $k_x$,  the behavior along $k_y$ is not affected by the new coupling and the Fermi surface is still present as $n$ is increased (see the right plot of Fig.\,\ref{fig:AvskPDW}). 
This is qualitatively analogous to the behavior seen in the explicit lattice cases studied in Section~\ref{explicit_case}. 
Thus, from this analysis we conclude that when the inhomogeneity effect is strong enough, Fermi surfaces can be suppressed even in the case of purely spontaneous symmetry breaking, leaving behind disconnected Fermi surface segments reminiscent of Fermi arcs. 

We stress that the behavior in this particular model is due \emph{entirely} to the strength of the spontaneous IR stripes.
Indeed, if the gradual disappearance of the Fermi surface were caused by the 
new interaction term
$\chi^2\overline{\zeta}\zeta$, and not by the strong IR modulation,  
the spectral density along the $k_y$ direction would also be suppressed
with increasing $n$. Since it is clear from the right plot of Fig.\,\ref{fig:AvskPDW} that it isn't, we conclude that the segmentation 
phenomenon is inherently due to the strong IR inhomogeneity. 
Also, we note that the interaction term 
causes the outer Fermi surface 
to shrink, while at the same time triggering and enhancing the inner secondary Fermi surface.
Such kind of behavior can be seen more clearly from the density plot of the spectral function momentum distribution for the CDW case, as we discuss below.

\subsubsection{Charge Density Wave Case}

\begin{figure}[ht!]
\begin{center}
\includegraphics[width=.49\textwidth]{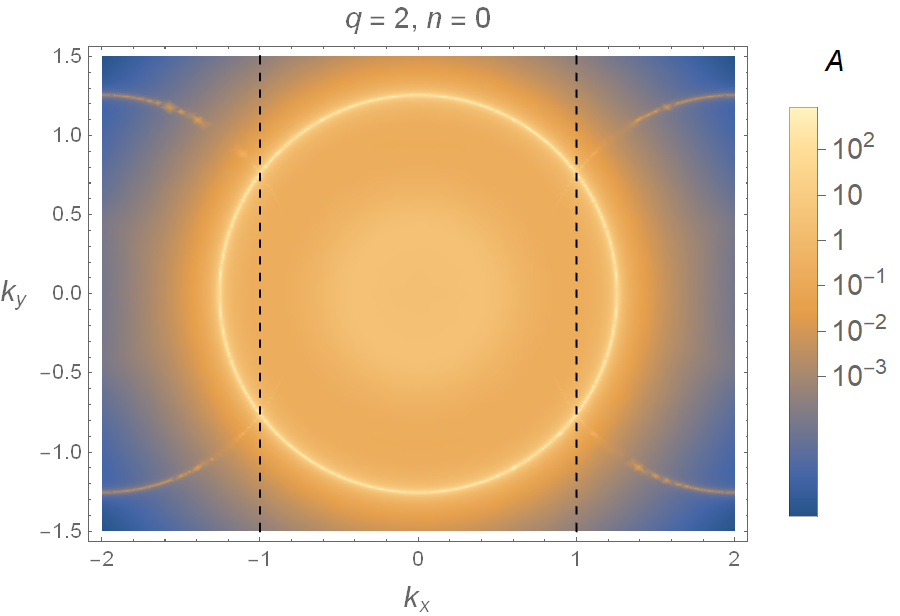}\;\;
\includegraphics[width=.49\textwidth]{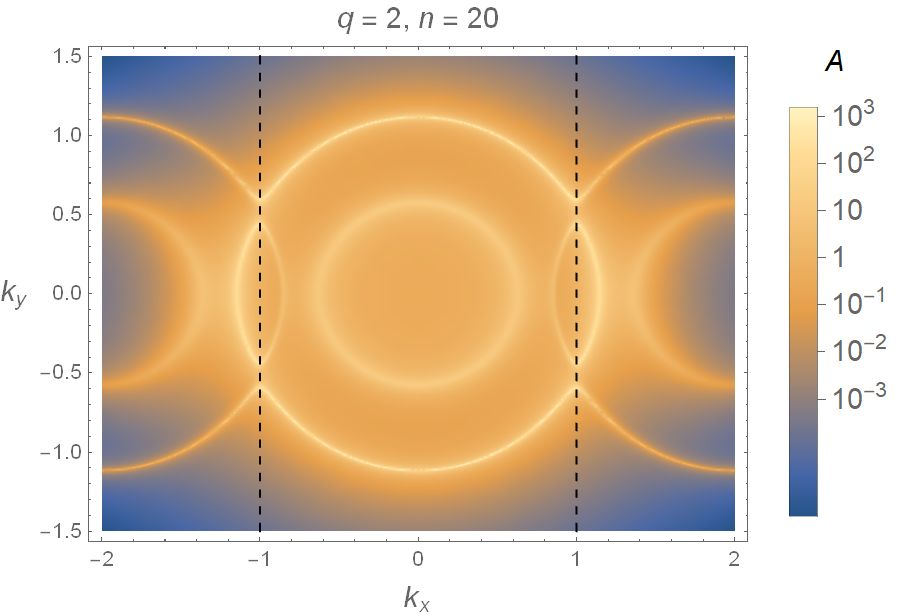}\\
\vspace{10pt}
\includegraphics[width=.49\textwidth]{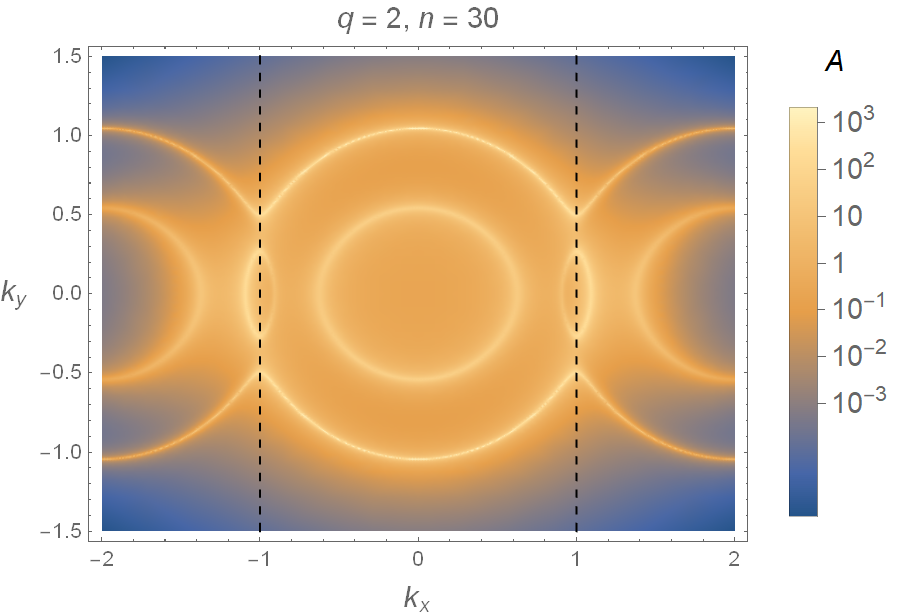}\;\;
\includegraphics[width=.49\textwidth]{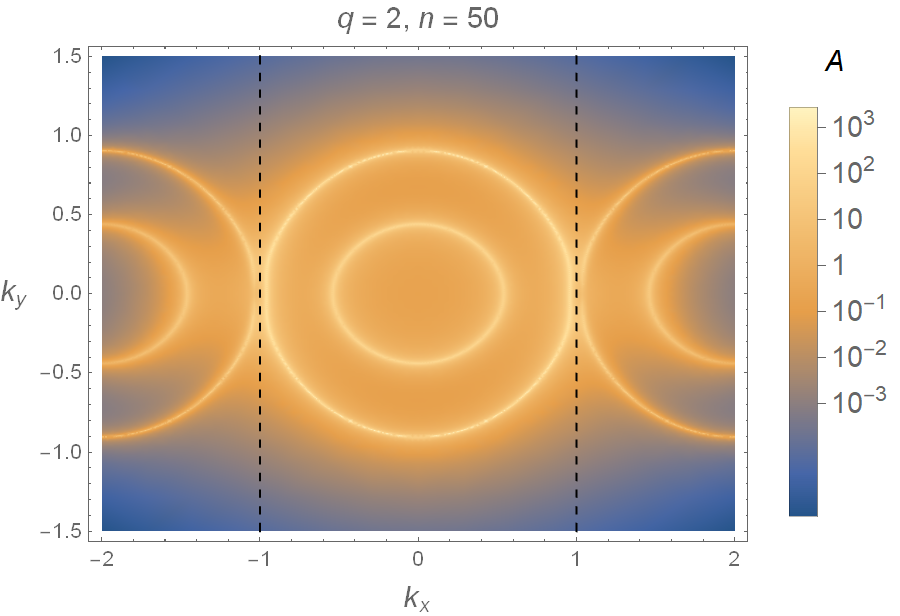}
\caption{The density plot of the momentum distribution of the spectral function in the $(k_x, k_y)$ plane for $q=2$ and $\omega=10^{-6}$. The brightest points correspond to the location of the Fermi surface. 
The first Brillouin zone boundary is denoted by the vertical dashed lines at $k_x=\pm 1$, and the CDW background geometry has $T=0.02144$ and $k=1$. The four plots correspond to the parameter $n=0$, $20$, $30$, and $50$. }
\label{fig:FS_CDW_q2}
\end{center}
\end{figure}

\begin{figure}[ht!]
\begin{center}
\includegraphics[width=.49\textwidth]{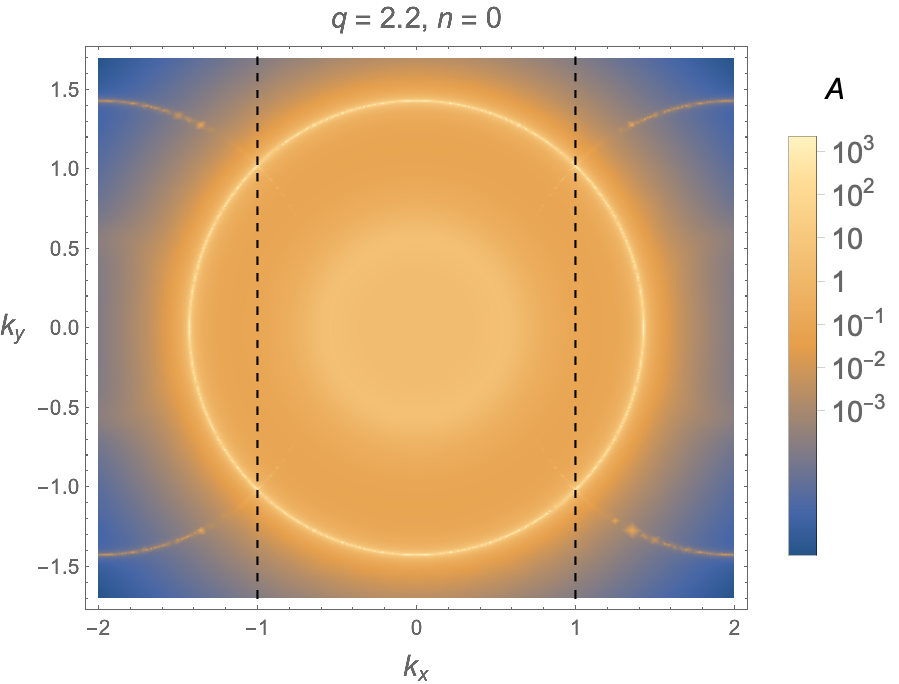}\;\;
\includegraphics[width=.49\textwidth]{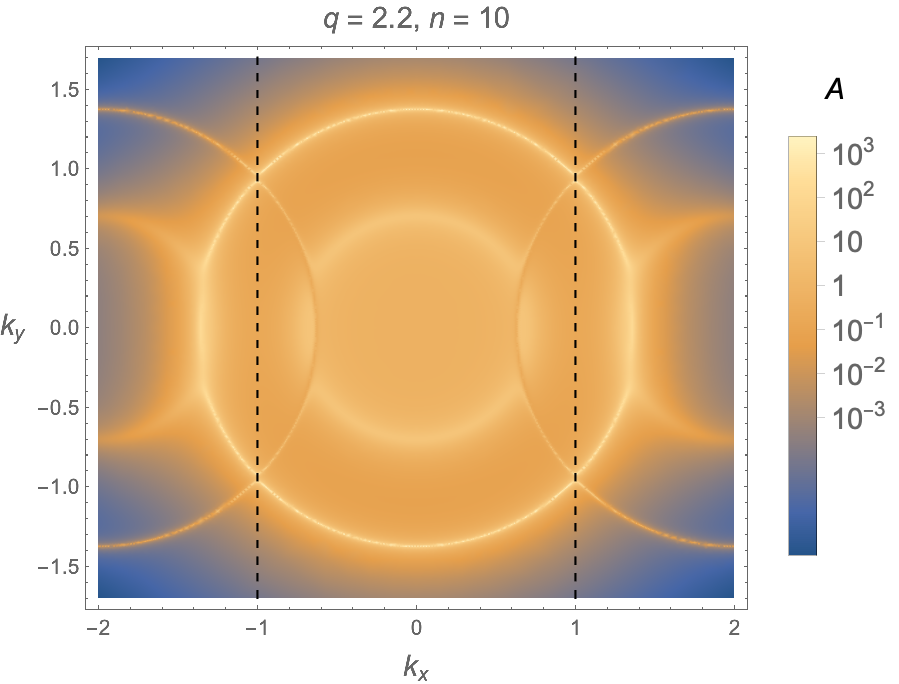}\\
\vspace{10pt}
\includegraphics[width=.49\textwidth]{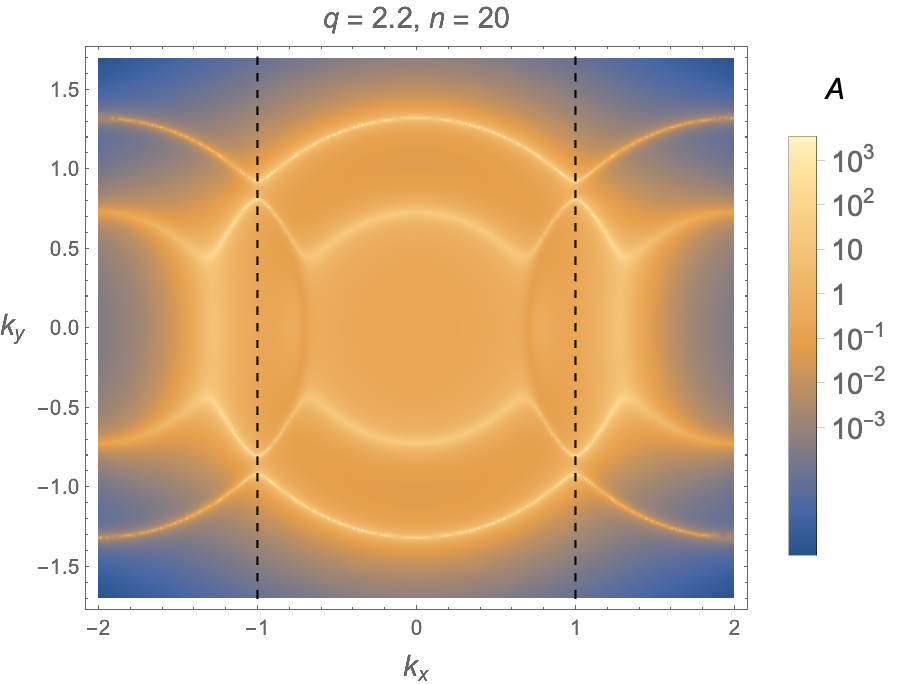}\;\;
\includegraphics[width=.49\textwidth]{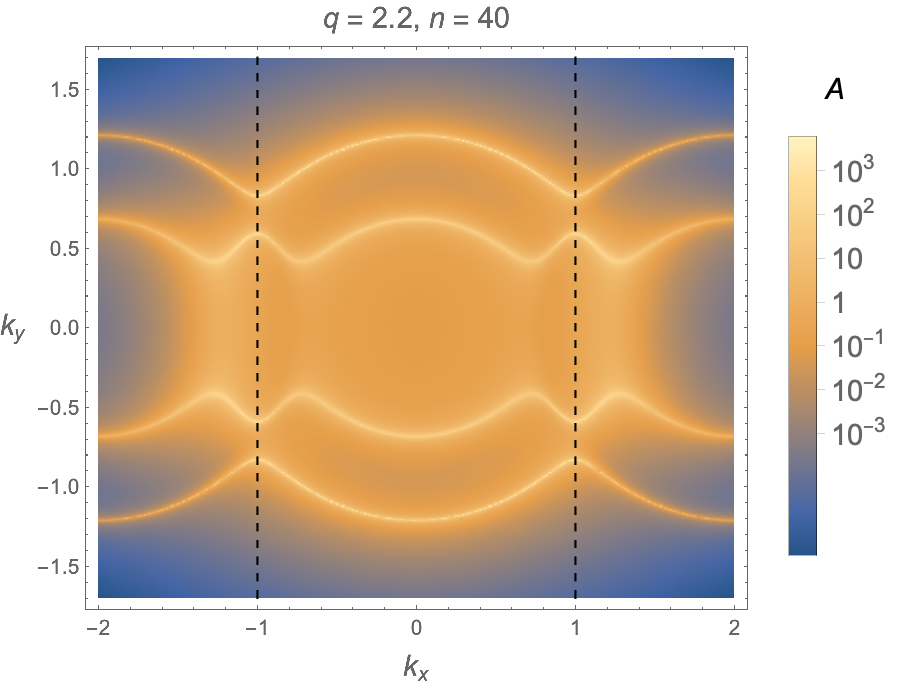}
\caption{The density plot of the momentum distribution of the spectral function in the $(k_x, k_y)$ plane for $q=2.2$ and $\omega=10^{-6}$. The brightest points correspond to the location of the Fermi surface. 
The first Brillouin zone boundary is denoted by the vertical dashed lines at $k_x=\pm 1$, and the CDW background geometry has $T=0.02144$ and $k=1$. The four plots correspond to the parameter $n=0$, $10$, $20$, and $40$. As $n$ is increased, some vertical parts of the Fermi surface gradually vanish.}
\label{fig:FS_CDW_q22}
\end{center}
\end{figure}

\begin{figure}[ht!]
\begin{center}
\includegraphics[width=.49\textwidth]{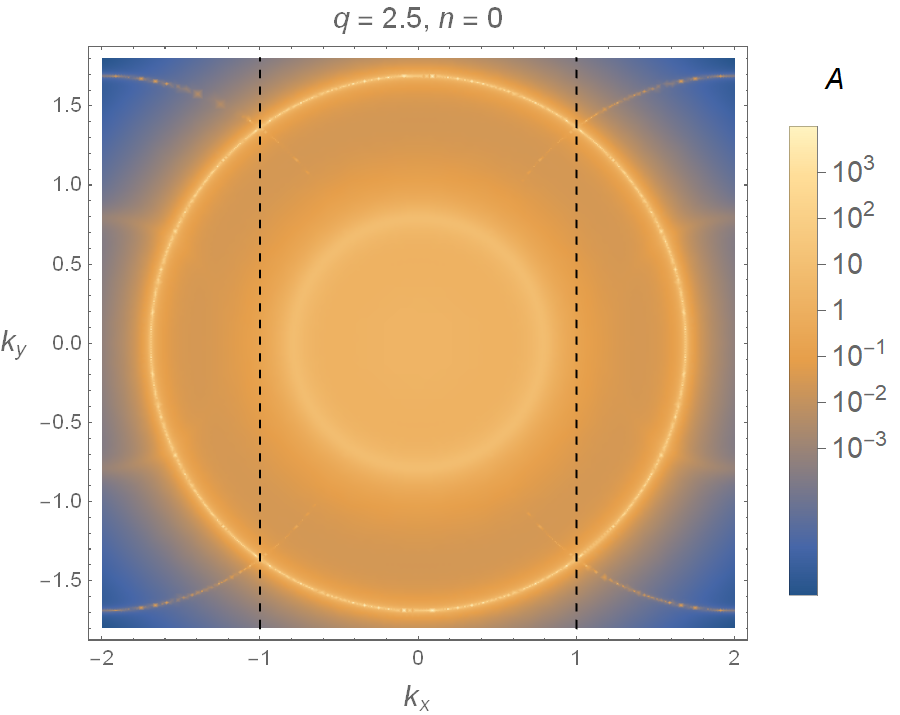}\;\;
\includegraphics[width=.49\textwidth]{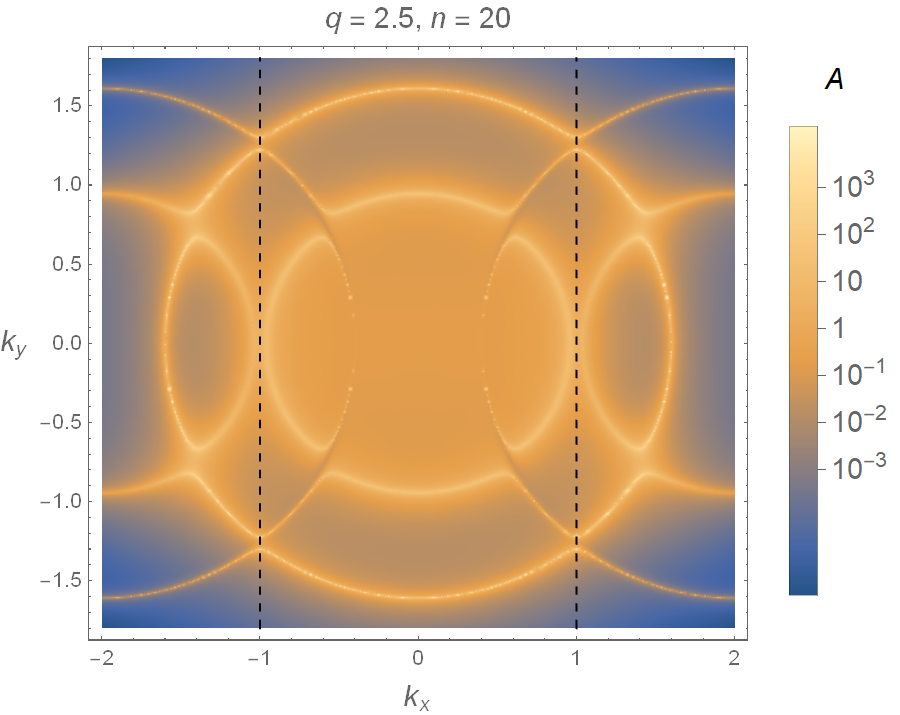}\\
\vspace{10pt}
\includegraphics[width=.49\textwidth]{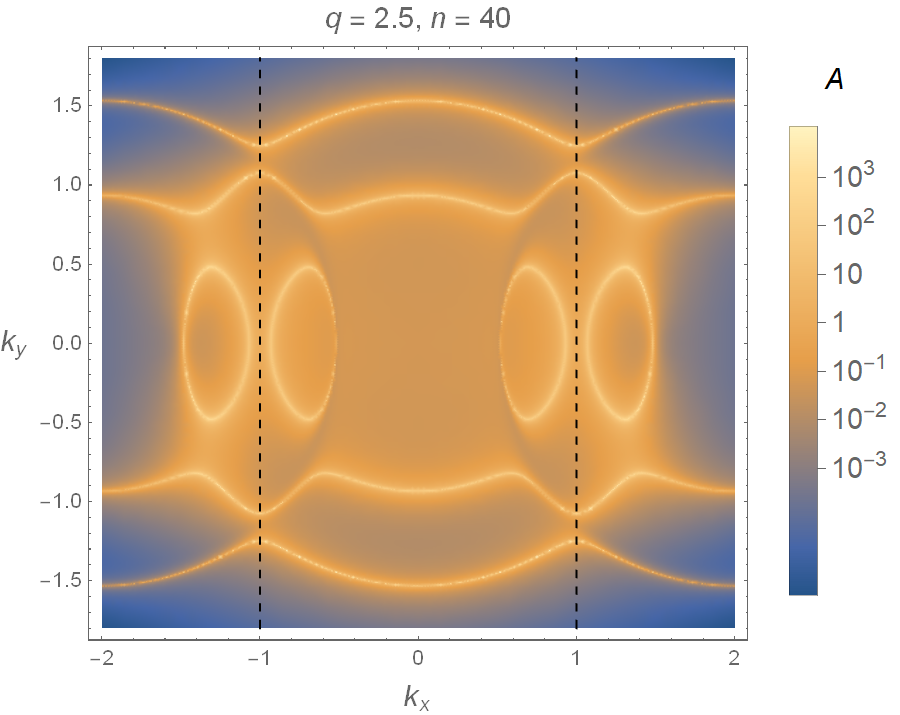}\;\;
\includegraphics[width=.49\textwidth]{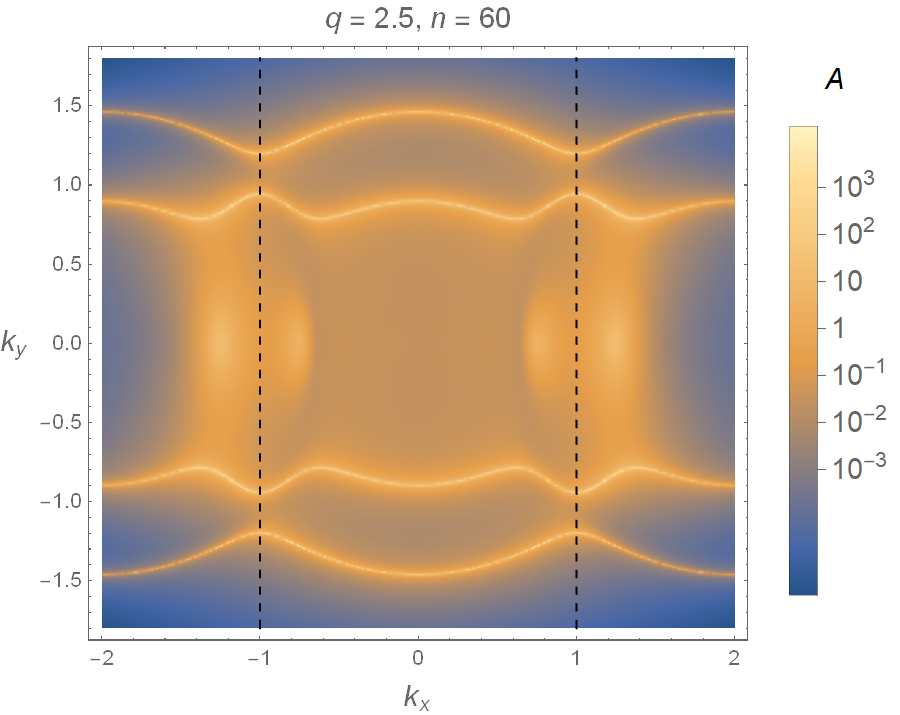}
\caption{The density plot of the momentum distribution of the spectral function in the $(k_x, k_y)$ plane for $q=2.5$ and $\omega=10^{-6}$. The brightest points correspond to the location of the Fermi surface. 
The first Brillouin zone boundary is denoted by the vertical dashed lines at $k_x=\pm 1$, and the CDW background geometry has $T=0.02144$ and $k=1$. The four plots correspond to the parameter $n=0$, $20$, $40$, and $60$. As $n$ is increased, some vertical parts of the Fermi surface gradually vanish.}
\label{fig:FS_CDW_q25}
\end{center}
\end{figure}

We now restrict our attention to a pure CDW state, in which the U(1) symmetry is not broken (we set $q_A=q_B=0$ and consistently truncate $\theta$). 
We take the couplings in~\eqref{generalmodel} to be given by
\begin{equation}
\label{CDWcouplings}
\begin{split}
& Z_A(\chi)=1+2\chi^2,\quad Z_B(\chi)=1,\quad Z_{AB}(\chi)=-2.15\, \chi\,, \\
& V( \chi)=-\frac{1}{L^2}\chi^2,\quad \mathcal{K}(\chi)=0\,.
\end{split}
\end{equation}
The behavior of the Fermi surface as the strength of the inhomogeneity is increased is shown in Figs.\,\ref{fig:FS_CDW_q2}, \ref{fig:FS_CDW_q22} and \ref{fig:FS_CDW_q25} for,
respectively, $q=2$, $q= 2.2$ and $q=2.5$. In all of these plots we have used the spectral density representation (\ref{newspectral}). 
It is immediately clear that there is an asymmetry across the Brillouin zone boundary.
In particular, we identify what looks like a pocket near the zone boundary (\emph{e.g.} see the $n=20$ and $n=30$ cases 
of Fig.\,\ref{fig:FS_CDW_q2}), with 
one half of the pocket clearly brighter than the other half. 
This asymmetry is  reminiscent of the ARPES Fermi arc experimental measurements, which also 
display a similar structure across the zone boundary. 

\begin{figure}[ht!]
\begin{center}
\includegraphics[width=.48\textwidth]{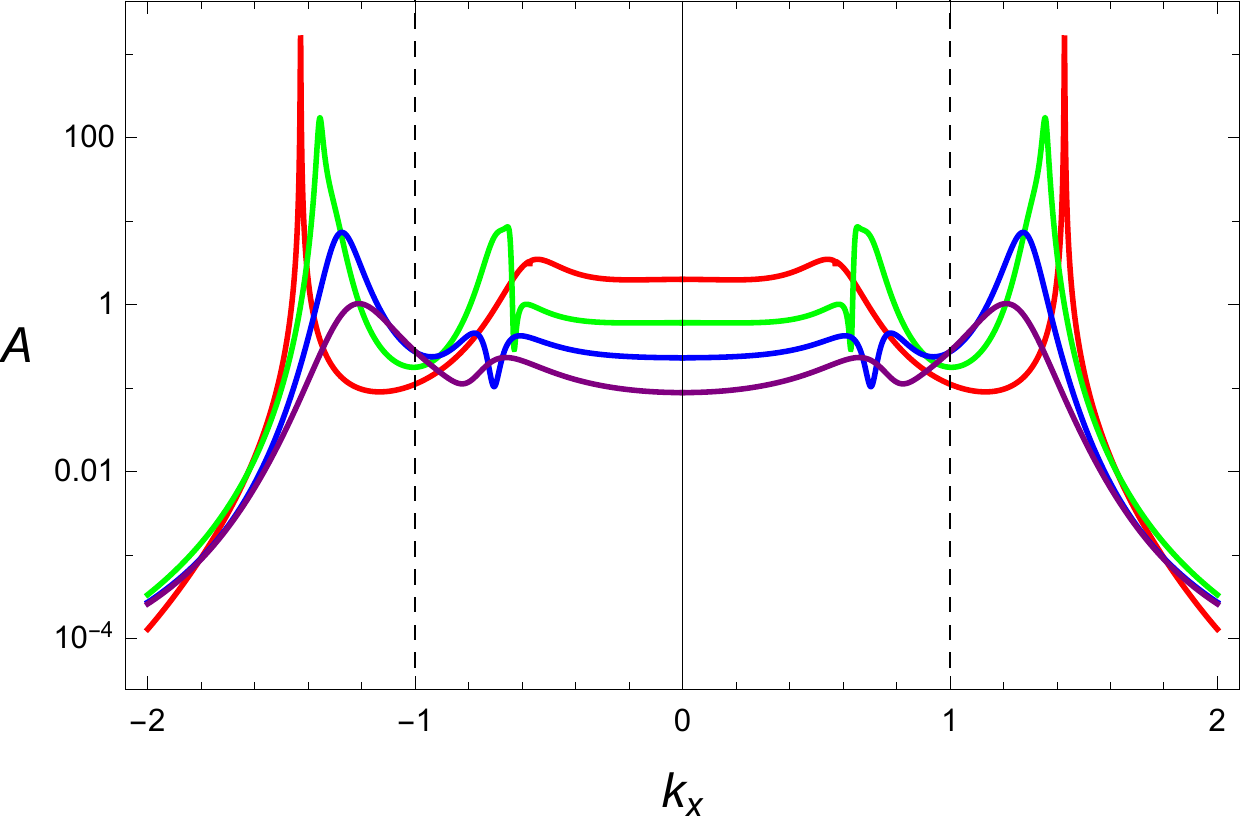}\quad
\includegraphics[width=.48\textwidth]{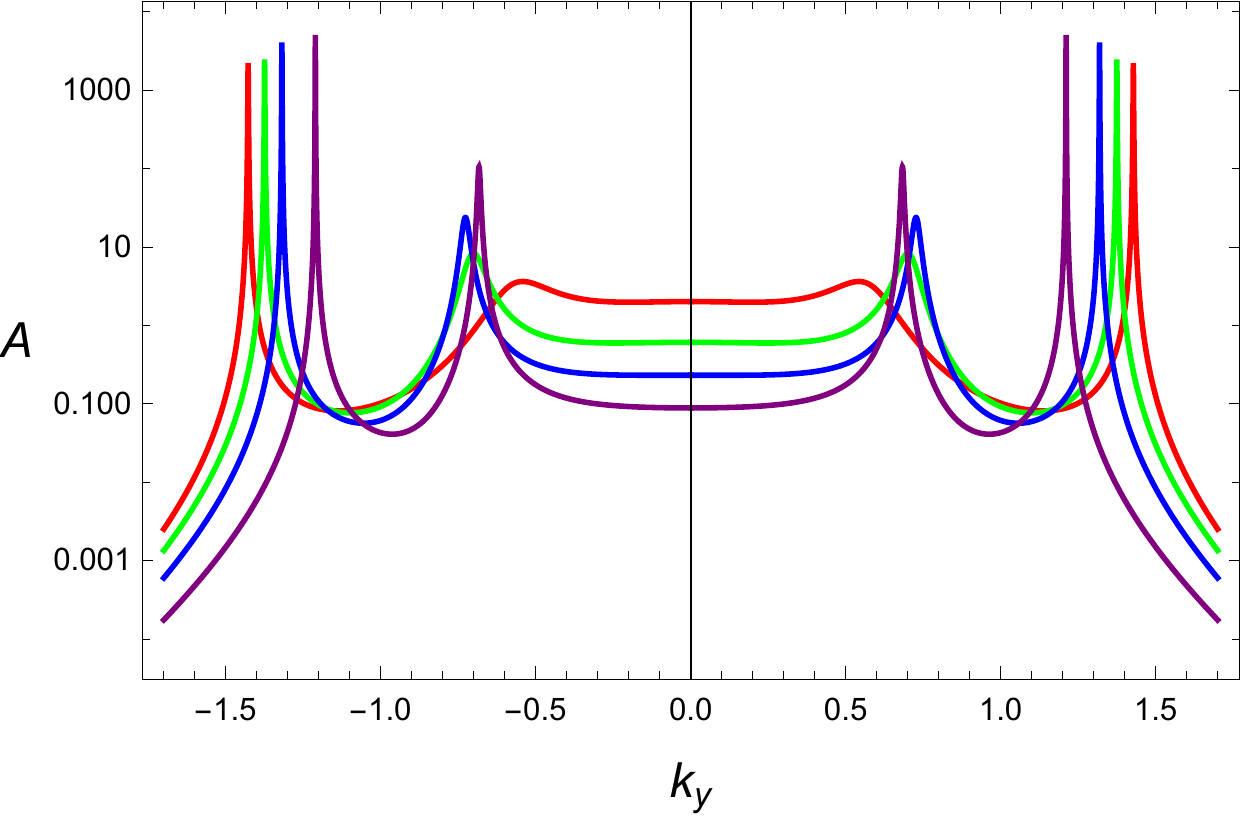}
\caption{Momentum distribution of the spectral density along the $k_x$ axis (left panel) for $k_y=0$ and along the $k_y$ axis (right panel) for $k_x=0$ for varying values of the coupling constant $n$. 
Translational invariance is broken spontaneously along the $x$ direction and the first Brillouin zone boundary is at $k_x=1$. In both plots the red curve corresponds to $n=0$, while the green, blue and purple curves correspond to $n=10, 20$ and $40$, respectively. We have fixed $\omega=10^{-6}$, $q=2.2$ and chosen $T=0.02144$ for the CDW geometry. }
\label{fig:AvskCDW}
\end{center}
\end{figure}

As one can see, the band gap at the Brillouin zone boundaries grows with the size of the parameter $n$. 
Moreover, increasing the effect of the IR spontaneous modulation does cause a gradual disappearance of the Fermi surface along the symmetry breaking direction, eventually leading to the formation of small disconnected Fermi surface segments\,\footnote{Note, however, that when $n=50$ the $q=2$ Fermi surface in 
Fig.\,\ref{fig:FS_CDW_q2} is almost entirely contained inside the first Brillouin zone, and appears to be completely closed, as expected from the discussion of \ref{ioniclatticesubsection}.}.
This trend is particularly clear from the last two plots of Fig.\,\ref{fig:FS_CDW_q22}.
For this value of the charge ($q=2.2$) the behavior of the ``secondary Fermi surface,'' the blurry bump visible in Fig.~\ref{fig:FS_CDW_q22}, is more striking. 
As $n$ is increased, the inner bump  becomes larger and sharper along the $k_y$ direction, while it is suppressed along the $k_x$ direction. 
This behavior is clearly visible in the corresponding momentum distribution of the spectral density along the $k_x$ and $k_y$ directions, which is given in Fig.\,\ref{fig:AvskCDW}.
In addition to the segmentation of the Fermi surface being generic with strong translational symmetry breaking,
increasing the strength of the interaction term also appears to enhance the (inner) secondary Fermi surface.
However, the detailed behavior of the latter seems to be somewhat model dependent. 
For instance, the structure of the Fermi surface in Fig.\,\ref{fig:FS_CDW_q25} corresponding to $q=2.5$ is even more elaborate 
than in the previous cases.
The trend common to all the examples is the gradual disappearance of some of the vertical portions of the Fermi surface, which is visible here as well.

%


From the cases we have studied here and in~\cite{Cremonini:2018xgj}
we conclude that, while 
the fine structure of the Fermi surface is very sensitive to the details of the theory (the detailed 
dependence on the charge $q$ and on the mass term $n$ is rather complicated), 
in strongly correlated systems Fermi surfaces can be generically suppressed when the inhomogeneity effect is strong enough. 
Since we also observe a similar spectral density suppression in the CDW case without  the U(1) symmetry breaking, the broken U(1) doesn't play a key role in the process.
Despite the strong numerical evidence, however, 
the real origin of the spectral weight suppression is still unclear. 

Our analysis also shows that strong lattice effects lead to a segmentation and 
gradual destruction of the ``Fermi pocket'' near the Brillouin zone and along the $k_x$ axis -- it is clear from our numerics 
that there is no sharp spectral density near $\omega=0$.
There are different possible explanations for this behavior.  
Since the Fermi surface, by definition, lies at the intersection between the dispersion
relation $\omega(k_x, k_y)$ and the Fermi level $\omega=0$ (measured relative to the chemical potential), it is possible that increasing the lattice amplitude might lift the energy band above the Fermi level, due to the strong Umklapp eigenvalue repulsion\,\footnote{We are grateful to Alexander Krikun and Koenraad Schalm for raising this point with us.}. 
This could cause the Fermi pockets to become smaller and eventually disappear.
On the other hand, the poles of the Green's function in the presence of a strong lattice might  move to the complex-$\omega$ plane. 
If this were the case at $\omega=0$ one would only see a residual bump from the spectral density 
-- see, for example, the last plot of Fig.\,\ref{fig:FS_CDW_q25}.
More work is needed to isolate the different possible contributions to the process of the broadening of the peaks.
 Preliminary results on the energy distribution function, i.e. the spectral density as a function of $\omega$, are presented in Appendix~\ref{app:energy}. 
As we will see, there seems to be an energy gap opening near $\omega=0$ due to strong lattice effects.

\section{Summary of Results and Discussion}
\label{SectionDiscussion}

In this paper we have extended the analysis of our earlier work \cite{Cremonini:2018xgj} to several holographic models of lattices and (striped) phases with broken translational invariance. 
We have constructed the fermionic spectral function $A(\omega,\textbf{k})$ 
and adopted a working definition \cite{Cosnier-Horeau:2014qya} of the Fermi surface as being identified (at finite 
temperature) by sufficiently sharp peaks indicative of gapless excitations and
non-analytic behavior at $\textbf{k} = \textbf{k}_F$ and $\omega = 0$.
We have focused on examining its dependence on the fermionic charge and on the strength of lattice effects (or the strength of stripe order).
We summarize our analysis and main results here:
\begin{itemize}
\item
A Fermi surface forms when the charge $q$ of the fermion is sufficiently large. 
For systems with translational symmetry broken along one of the spatial directions (we have focused on uni-directional stripes oriented along the $x$ direction) 
the Fermi surface is anisotropic and, when it is larger than the first Brillouin zone, 
it typically displays a band gap\,\footnote{Note that this is not a superconducting gap -- it is due to the broken translational invariance.}
 at the zone boundary, due to standard eigenvalue repulsion.
The opening of the gap 
was observed early in the holographic literature~\cite{Liu:2012tr,Ling:2013aya}, and more recently in~\cite{Cremonini:2018xgj}. Additional Fermi surfaces are expected to form once the charge is increased further, resulting in patterns of various complexity (see~\cite{Gubser:2012yb} for an analytical study exhibiting multiple Fermi surfaces).
Introducing the effective mass term $M(\chi) \sim n\chi^2$ also appears to trigger new Fermi surfaces.

\item
In every example we have studied, the fermionic spectral function is suppressed when the symmetry breaking becomes sufficiently strong, independently of 
whether translational invariance is broken explicitly or spontaneously. 
In particular, unlike in~\cite{Cremonini:2018xgj} we have now constructed a model where the suppression is present when the symmetry breaking is purely spontaneous (we have done so by coupling the fermionic field to a scalar through an interaction term that acts as an effective mass for the fermion).
Thus, our analysis indicates that the suppression of the fermionic spectral weight with strong lattice effects is a universal feature, confirming and generalizing the results of~\cite{Cremonini:2018xgj}. However, this phenomenon happens only when the Fermi surface extends beyond the first Brillouin zone\,\footnote{This seems quite natural, since when the Brillouin zone is sufficiently large as compared to the Fermi momentum, the Fermi surface is entirely within the first Brillouin zone and therefore the effect of the lattice on the fermion response is small. As a consequence,  the Fermi surface should be almost isotropic, as confirmed by Fig.\,\ref{fig:Avskionicq20k36}.}.
Indeed, this indicates that the suppression is a result of some kind of interaction between Fermi surface branches belonging to different zones.

\item 
We have computed the spectral function using two different definitions. 
The first, given in (\ref{newspectral}) and used throughout the main text of the paper, does not implement any folding procedure and can be thought of as the analog of the extended scheme. By using (\ref{newspectral}) we retain the asymmetric properties of the Fermi surface across the zone boundaries, thus facilitating the comparison with ARPES measurements, which directly probe the 
electronic band structure in the extended zone.  
In particular, notice that in the density plot of Fig.\,\ref{fig:FS_CDW_q2} one can clearly see a feature similar to a Fermi arc.

The second definition, given in (\ref{spectral}) and adopted in the Appendix, ensures that the branches outside the first Brillouin zone are folded back into it and 
provides a representation of the band structure in the reduced zone scheme. The resulting Fermi surface is by construction 
automatically symmetric across the Brillouin zone boundary. 
Interestingly, the spectral function in the extended zone scheme in our model is not necessarily periodic, as discussed already in the earlier work~\cite{Liu:2012tr}, which argued that such non-periodicity 
is expected for non-Fermi liquid. 
In particular, the lack of periodicity is more visible when lattice effects are strong. As a side effect, 
any sharp spectral density peak associated with the Fermi surface is~\emph{enhanced} in the folded representation (see 
the discussion in Appendix \ref{Appendix}). 
In turn this implies that the folded version (\ref{spectral}) makes it slightly easier to see if the spectral weight is indeed suppressed. 
Nonetheless, we stress that the suppression in the presence of 
strong inhomogeneity occurs independently of which spectral function representation we adopt.

\item
The cases that involve multiple Fermi surfaces are interesting in their own right and lead to more elaborate patterns.
This is visible for example in Figs. \ref{fig:FS_CDW_q22} and \ref{fig:FS_CDW_q25}, where 
a secondary Fermi surface forms as the lattice strength is increased.
In these examples the suppression of the fermionic spectral function with strong lattice effects still occurs, but its detailed structure is now complicated by the presence of the secondary Fermi surface
(see also the energy distribution of the spetral function, shown in Fig. \ref{fig:DISP_CDW_q22}). 
The latter appears to interact non-trivially with the first Fermi surface branch, leading to 
more complex patterns of spectral weight suppression.
The detailed role of the secondary Fermi surface does not appear to be generic, but rather model dependent and begs for further understanding. 
Still, we don't think that the appearance of a secondary Fermi surface plays a fundamental (model-independent) role in the spectral weight suppression process 
(see also the discussion in \cite{Cremonini:2018xgj},
where we provided examples in which the suppression clearly occurred even when the secondary Fermi surface peaks were very well separated from the main Fermi surface peaks). 
\end{itemize}

Many open questions and challenges remain.
While from our analysis we can conclude that the spectral weight suppression with strong lattice effects appears to be  
generic\,\footnote{This suppression could be a universal property of any correlator -- 
when different modes interact and momentum conservation is strongly relaxed, one can expect that all quasiparticles decay fast. We thank Elias Kiritsis for bringing up this point.}, we still don't understand its basic origin.
Our results indicate that it is related to the interplay between Fermi surface branches belonging to different Brillouin zones -- 
indeed, the effect disappears when the Fermi surface lies entirely within the first Brillouin zone. 
Precisely how the interaction between different branches leads to the segmentation of the Fermi surface is an important 
process we still don't understand. 
The appearance of secondary Fermi surfaces adds an additional complication to this story and begs for further study, but
at this stage does not seem to lead to universal features. 
Moreover, preliminary data on the energy distribution of the spectral function shown in Appendix \ref{app:energy} (right panel of Fig.  \ref{fig:DISP_CDW_q22}) 
displays the opening of an energy gap  when 
the lattice is sufficiently strong, which appears to be due to a combination between a shift of the energy band and the presence of multiple bands.

As we have seen, many factors contribute to the process of the Fermi surface segmentation and disintegration, 
and there may be competing mechanisms.
Thus, it is important to disentangle the role played by each one, and to identify more precisely which features are inherent to strong coupling or tied to the breaking of translational invariance, as well as to the existence of multiple Fermi surfaces interfering with each other in a non-trivial way. 
Moreover, we would like to better understand how our results complement -- and fit into -- 
 what is currently known about the behavior of Fermi surfaces in strongly correlated systems in condensed matter, with the Fermi arcs seen in high temperature superconductors providing a natural starting point. 
It would also be interesting to extend our analysis to the case with two periodic lattices that are incommensurate with each other\,\footnote{The holographic description of the commensurability effect can be found, for example, in~\cite{Andrade:2017leb,Krikun:2017cyw}.},
and ask if the incommensurability leads to any qualitatively new features~\cite{Bak1982}.
Finally, it would be desirable to identify the quasi-normal mode spectrum in these systems, and better understand
the nature of the low-energy excitations and their dispersion.  In particular, it would be interesting to check if the spectral function exhibits a distinctive ``pole-zero" line shape, which has been argued to be a very general feature of quantum systems containing the continuum as a subpart~\cite{Gnezdilov:2018qdu}.
We leave these questions to future work.

\begin{acknowledgments}
We are grateful to Chinedu Ekuma, Eduardo Fradkin, Alexander Krikun, Koenraad Schalm, Elias Kiritsis and Jan Zaanen for comments on the draft. We also thank Roberto Emparan, David Huse, Subir Sachdev, Ariel Sommer  and Kai Sun
for insightful conversations.
S.C. is supported in part by the National Science Foundation Grant PHY-1620169.
L.L. is supported by the Chinese Academy of Sciences (CAS) Hundred-Talent Program.
J.R. is supported by a startup grant at Sun Yat-sen University under Grant No. 74130-18841203.

\end{acknowledgments}

\appendix
\renewcommand{\theequation}{\thesection.\arabic{equation}}
\addcontentsline{toc}{section}{Appendix}

\newpage

\section{The Spectral Density With Folding} 
\label{Appendix}

In the main text we have shown the momentum distribution of the spectral density using the extend zone representation~\eqref{newspectral} without implementing any folding procedure.
In this appendix we give the corresponding spectral density in terms of the reduced zone representation~\eqref{spectral}, which takes into account the folding of the Fermi surface by summing the diagonal components of the Green's function.  Since the spectral density is an even function of $k_x$ and $k_y$, we show the spectral density for positive $k_x$ and $k_y$ from Figs.\,\ref{fig:fdAvskionicq20} to~\ref{fig:fdAvskPDW} as well as~Fig.\,\ref{fig:fdAvskCDW} without any loss of generality. For the density plots of the momentum distribution in the $(k_x, k_y)$ plane from Figs.\,\ref{fig:fdFS_CDW_q2} to~\ref{fig:fdFS_CDW_q25}, we periodically extend the data from the first Brillouin zone to the other ones. Thus, all plots in this section will display a perfect symmetry between different sides of the Brillouin zone.

For the case where the Fermi surface (or a sharp peak of the spectral density) is within the first Brillouin zone, the reduced zone representation~\eqref{spectral} gives very similar results. This can be seen by comparing Fig.\,\ref{fig:Avskionicq15} with Fig.\,\ref{fig:fdAvskionicq15}, as well as Fig.\,\ref{fig:Avskionicq20k36} with Fig.\,\ref{fig:fdAvskionicq20k36}. In contrast, when the Fermi surface crosses the first Brillouin zone, the branches outside of the latter are folded back into it in the reduced zone representation~\eqref{spectral}.  
Although the spectral density is not at all periodic in the extended zone, a particularly interesting point is that if the peaks appearing in each Brillouin zone in the extend zone representation~\eqref{newspectral} are sufficiently sharp,  after folding they will be located at the same $k_x$ in the reduced zone representation~\eqref{spectral}. 
But this ceases to be the case when the lattice effects become sufficiently strong. 
Therefore, any sharp spectral density peak associated with the Fermi surface turns out to be enhanced in the folded representation. 
In turn this tells us that using the reduced zone representation it is quite easy to see if the spectral weight will be suppressed by strong lattice effects.

Another reason for adopting the reduced zone representation~\eqref{spectral} is the valid concern that, by working only with the diagonal components of the Green's function, one would neglect  interactions between different Brillouin zones. 
Indeed, as the strength of the spatial modulation increases, so do the non-diagonal components of the Green's function. To diagonalize the system via a unitary transformation is quite non-trivial, since the system contains the full range of Bloch indices $n = 0, \pm1, \pm2, ....$. 
However, we note that the trace of a matrix is invariant under unitary transformations and implicitly contains the effects associated with the non-diagonal components in the original basis. 
Moreover, our main purpose in this paper is to examine whether the spectral density is suppressed or not. 
As we have shown in this paper, the spectral weight is suppressed when the inhomogeneity effect is strong enough, independently of which representation we have adopted.

\subsection{Ionic lattice case}

\begin{figure}[ht!]
\begin{center}
\includegraphics[width=.485\textwidth]{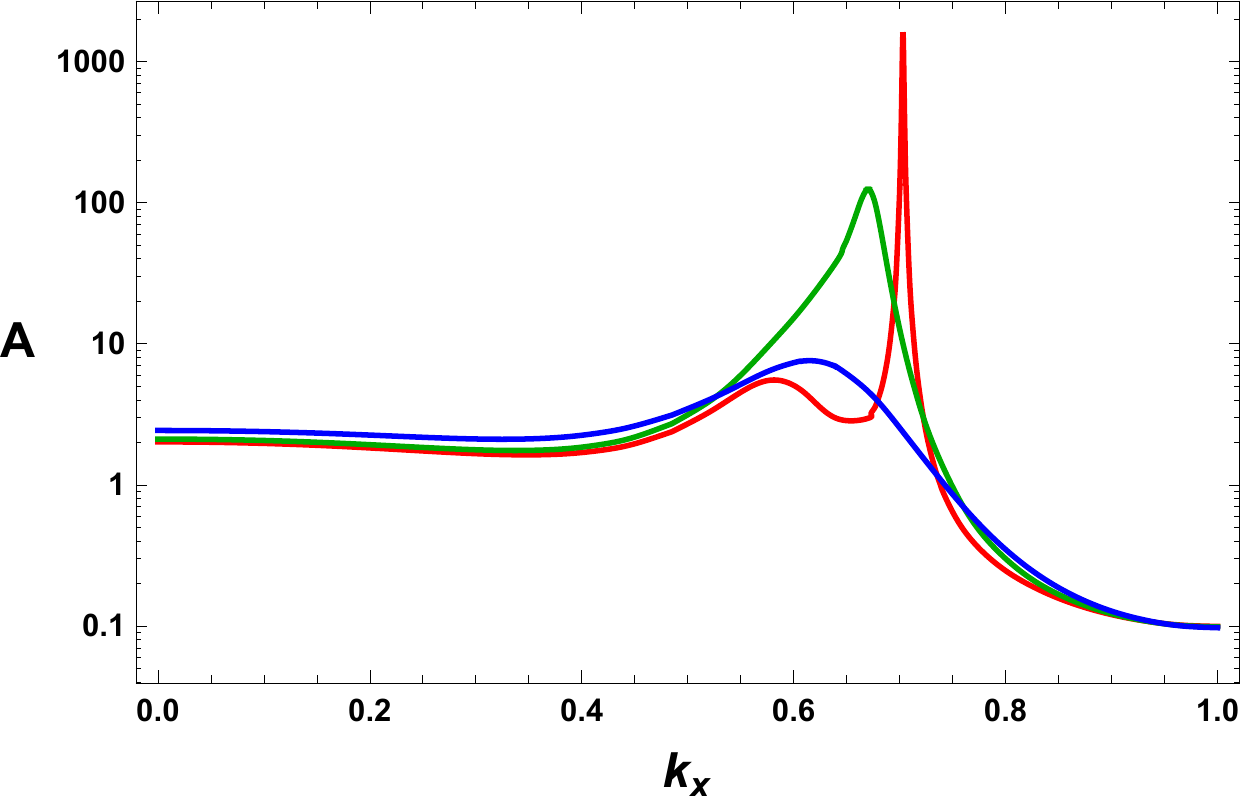}\quad
\includegraphics[width=.485\textwidth]{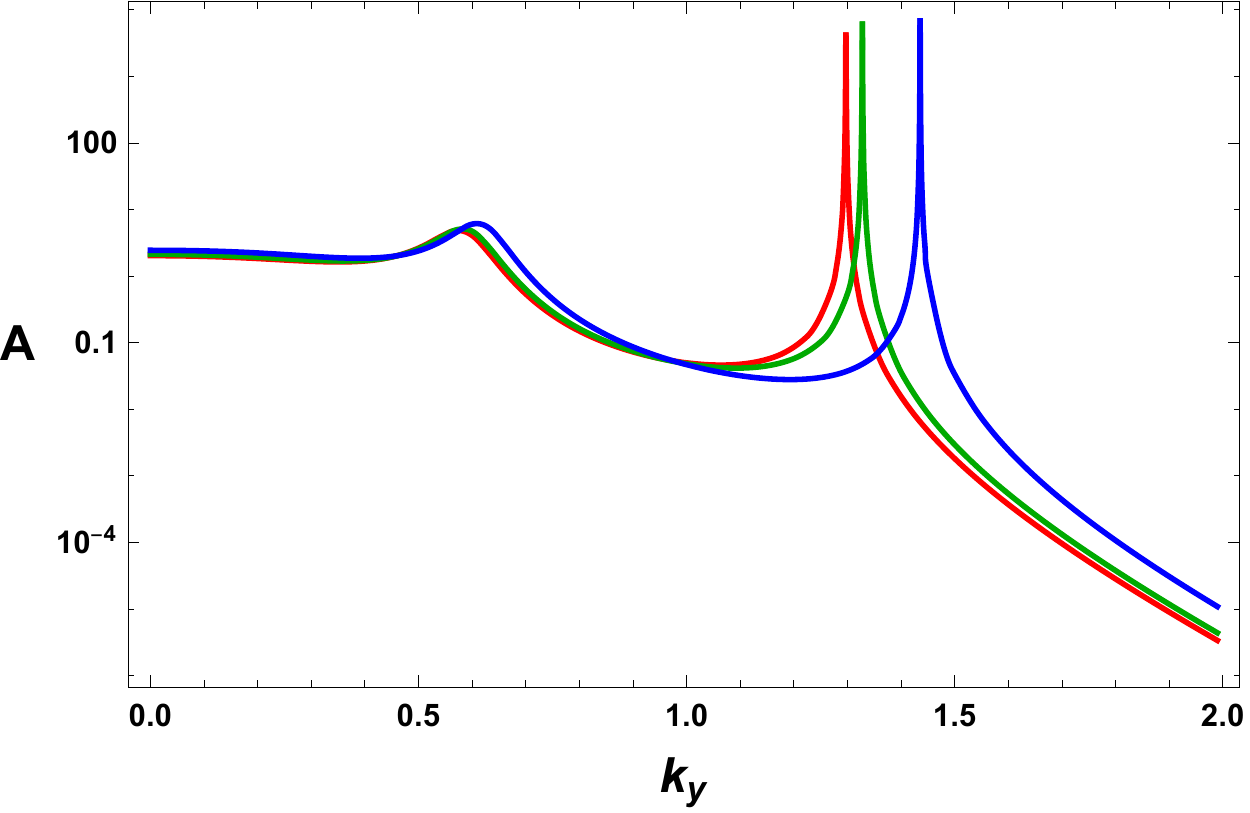}
\caption{Momentum distribution of the spectral density \eqref{spectral} along the $k_x$ axis (left panel) and the $k_y$ axis (right panel) for varying values of the amplitude of the ionic lattice. 
Translational invariance is broken explicitly in the $x$ direction and the first Brillouin zone boundary is at $k_x=1$.
The red, green and blue curves correspond to $a_0=0.1, 0.3$ and $0.6$, respectively. 
We have chosen $\omega=10^{-6}$, $q=2$, $T=0.0069$ and 
$p_I=2$. 
Since the vertical axis is logarithmic, the red curve describes a very sharp peak, indicative of a Fermi surface.}
\label{fig:fdAvskionicq20}
\end{center}
\end{figure}

\begin{figure}[ht!]
\begin{center}
\includegraphics[width=.485\textwidth]{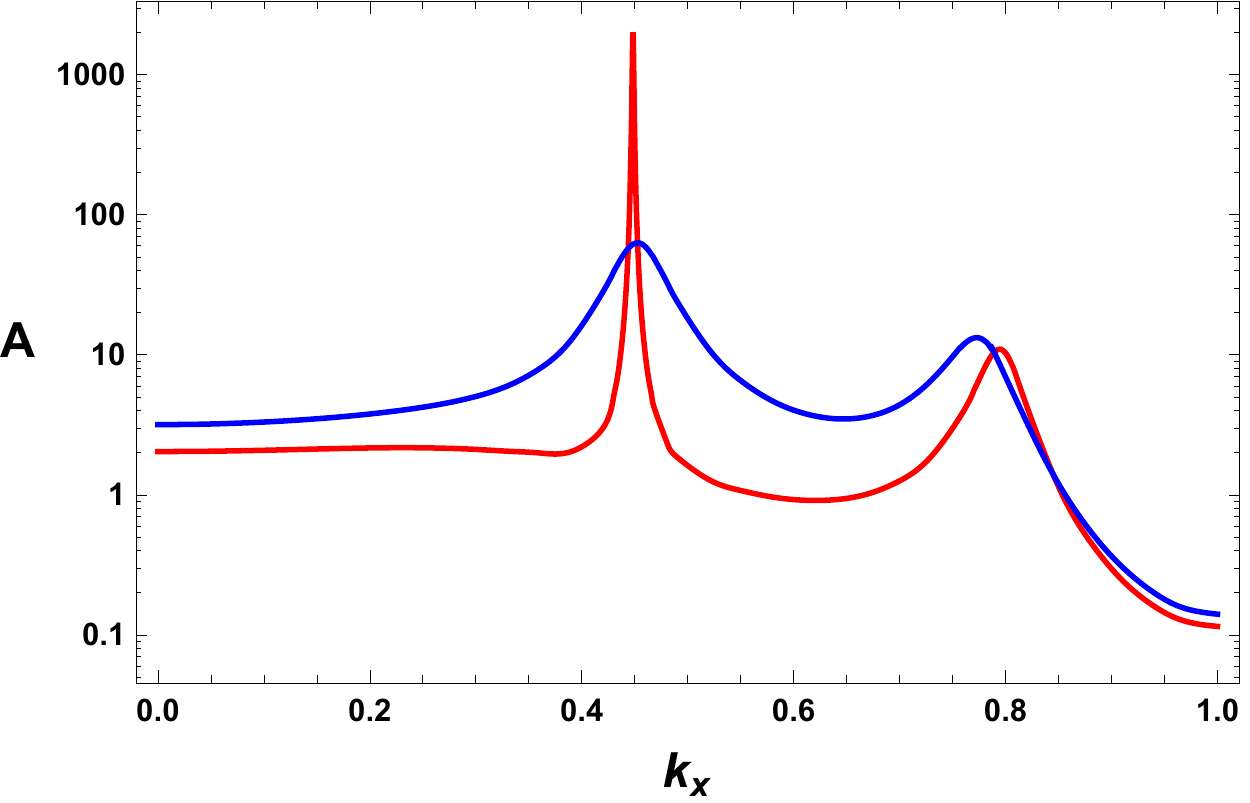}\quad
\includegraphics[width=.485\textwidth]{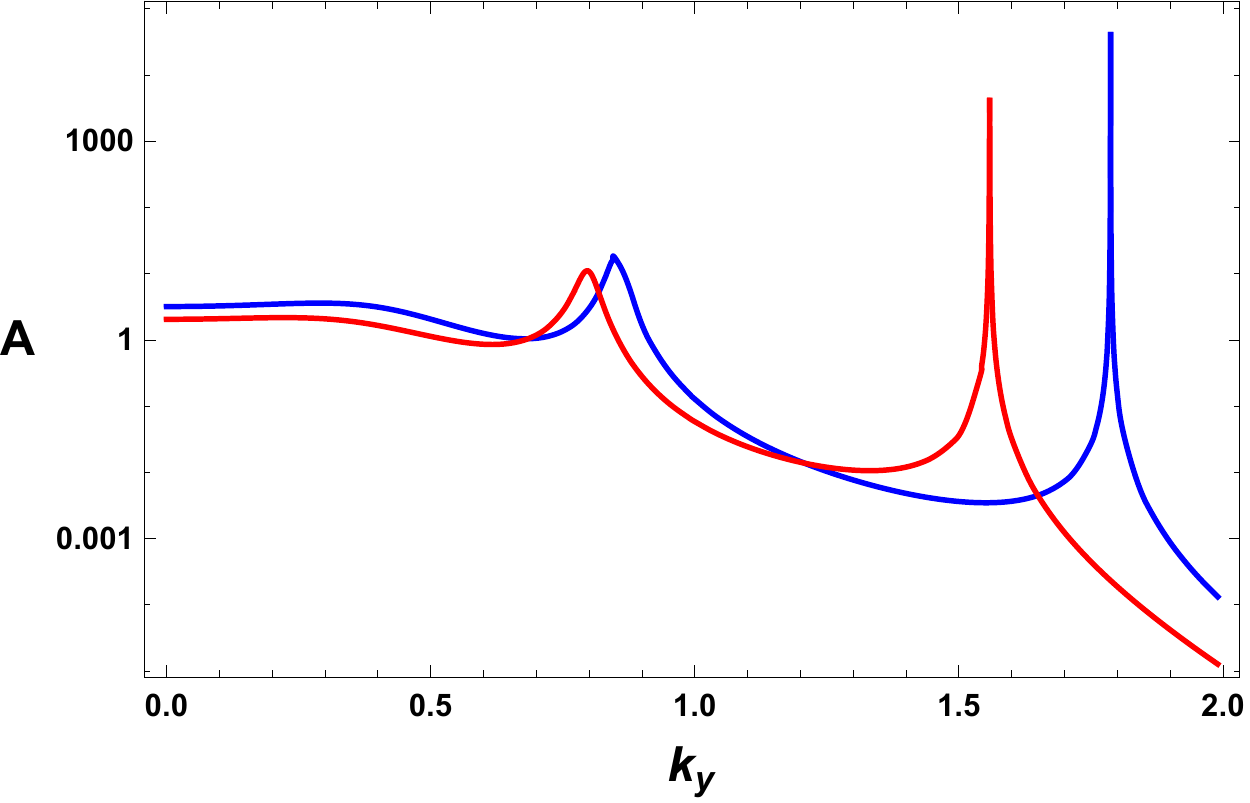}
\caption{Momentum distribution of the spectral density \eqref{spectral} for $q=2.3$ 
for lattice amplitudes $a_0=0.1$ (red curve) and $a_0=0.6$ (blue curve). 
The vertical axis is logarithmic. 
The remaining parameters are chosen as in Fig.\,\ref{fig:fdAvskionicq20}. 
}
\label{fig:fdAvskionicq23}
\end{center}
\end{figure}

\begin{figure}[ht!]
\begin{center}
\includegraphics[width=.485\textwidth]{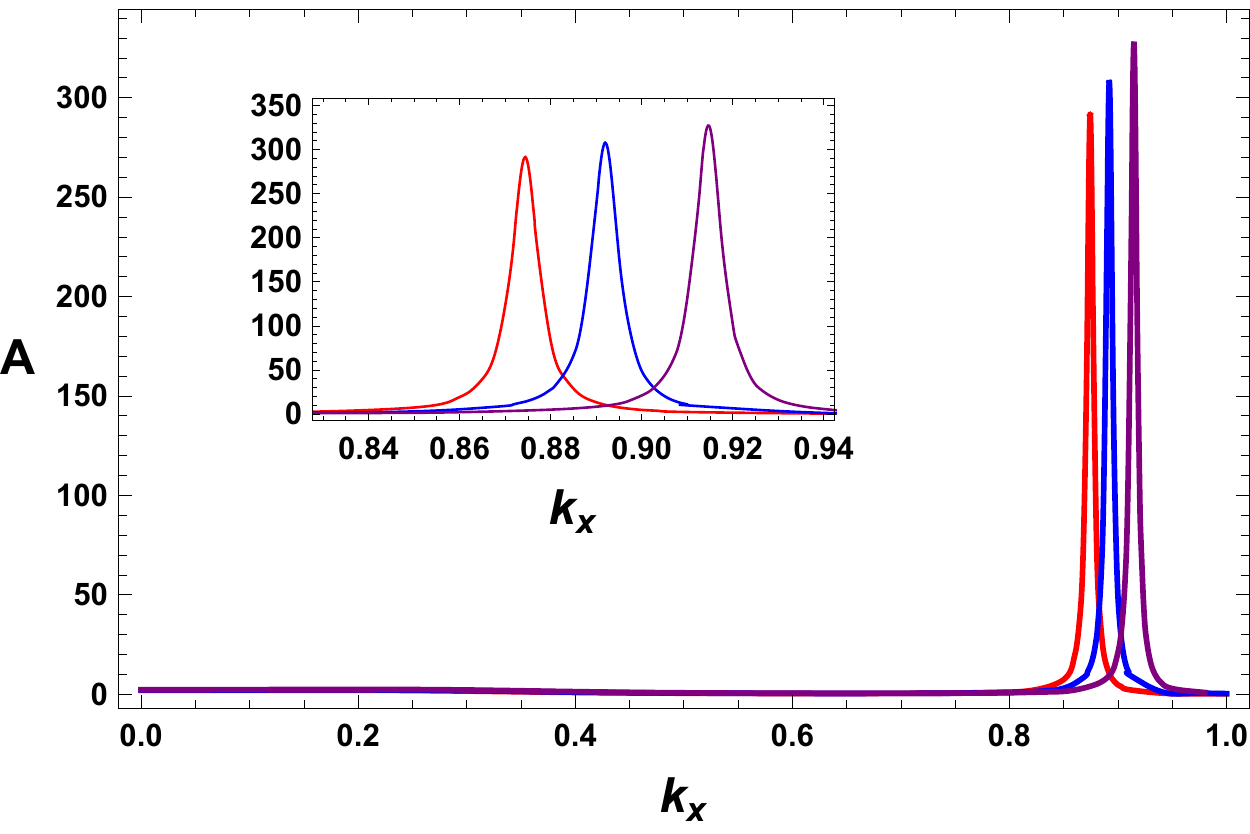}\quad
\includegraphics[width=.485\textwidth]{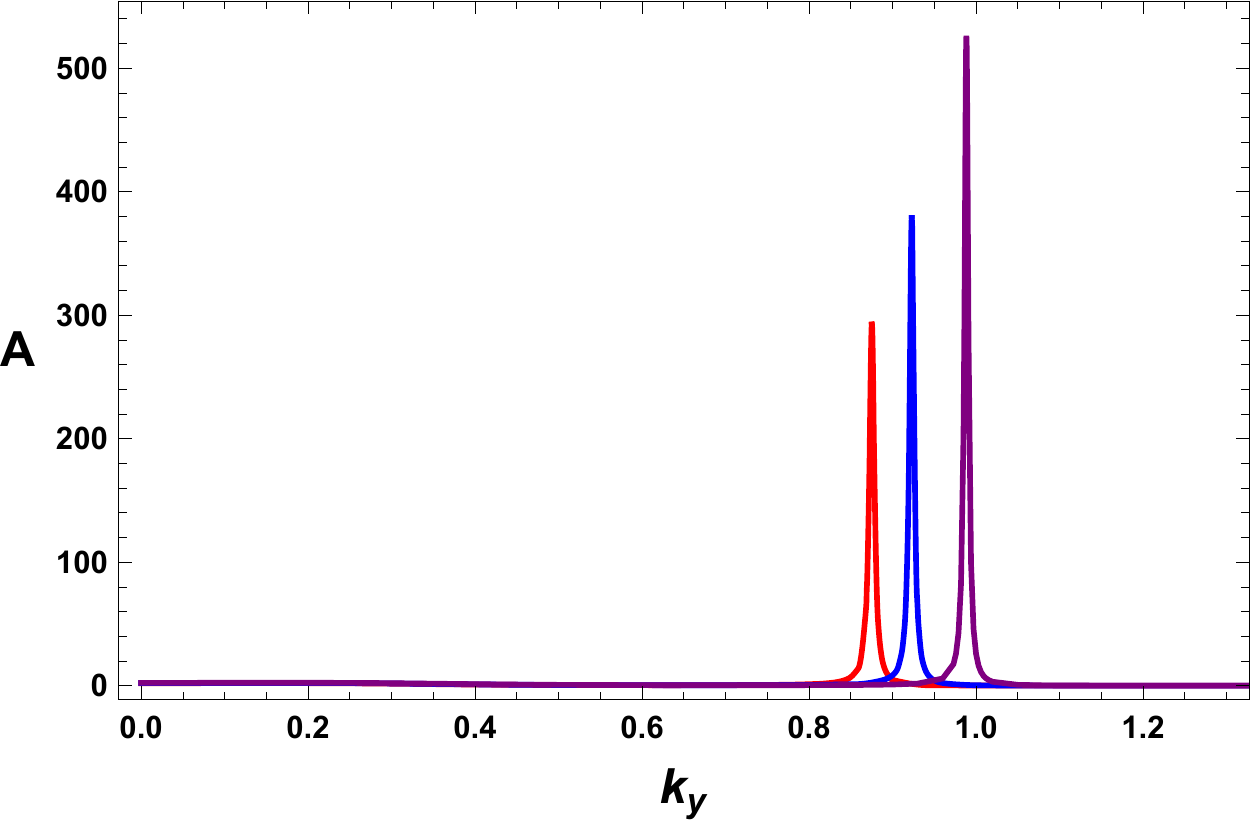}
\caption{Momentum distribution of the spectral density \eqref{spectral} for $q=1.5$ 
for ionic lattice amplitudes $a_0=0.1$ (red curve), $a_0=0.6$ (blue curve) and $a_0=0.9$ (purple curve). 
Here the vertical axis is not logarithmic.
All remaining parameters were chosen as in Fig.\,\ref{fig:fdAvskionicq20}. The inset in the left panel zooms into the location of the three peaks.
}
\label{fig:fdAvskionicq15}
\end{center}
\end{figure}

\begin{figure}[ht!]
\begin{center}
\includegraphics[width=.485\textwidth]{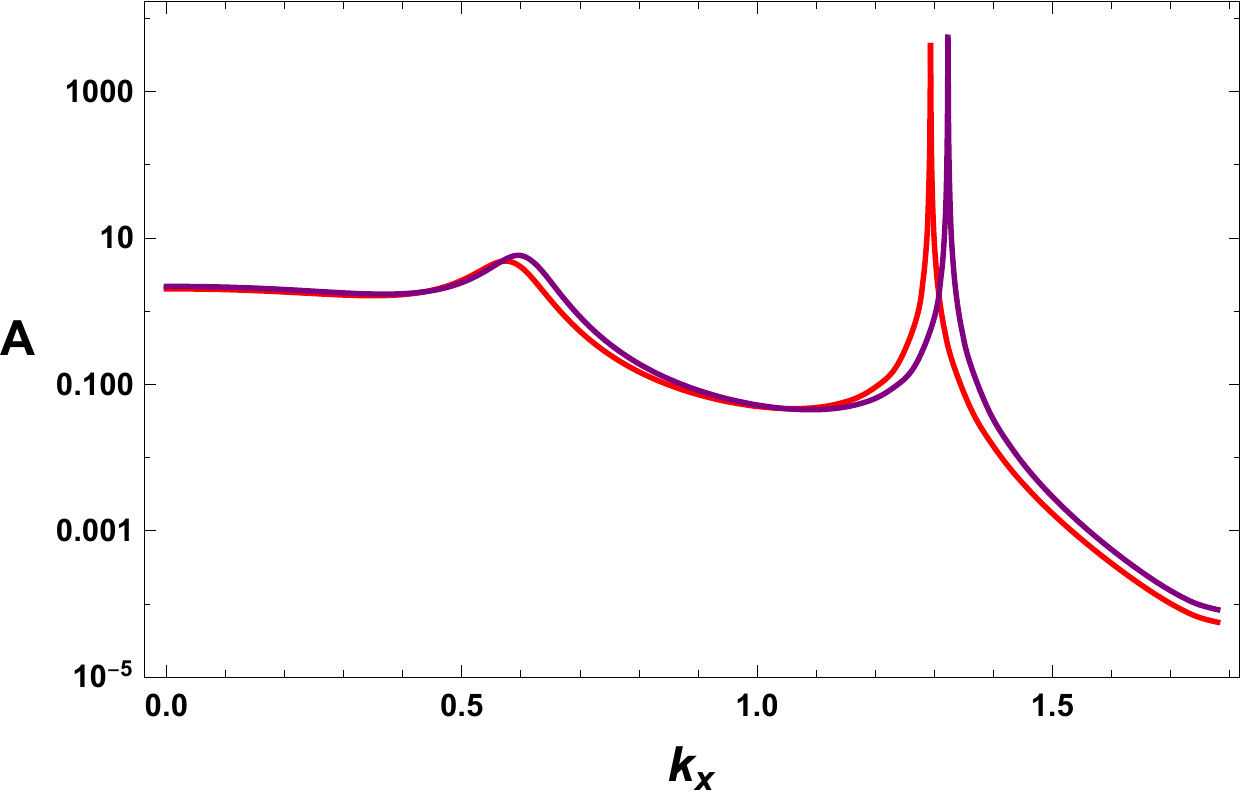}\quad
\includegraphics[width=.485\textwidth]{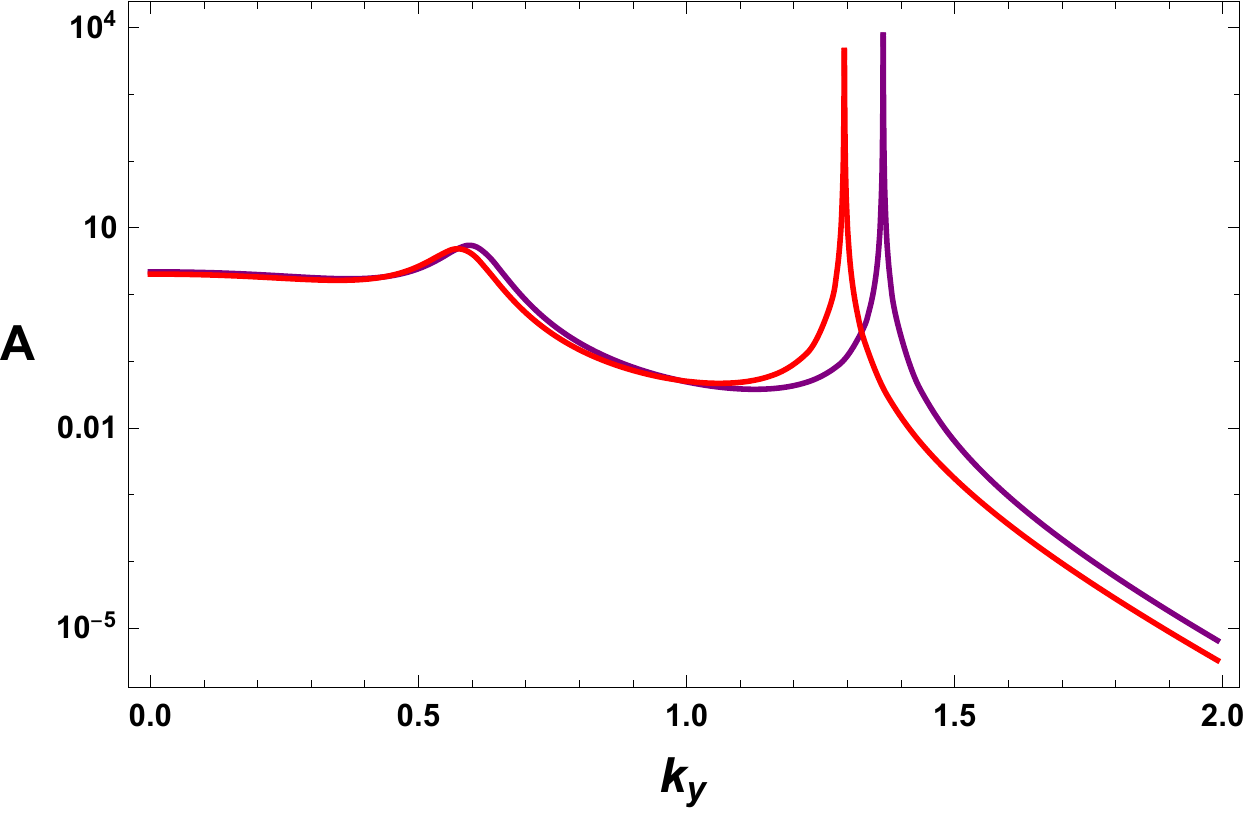}
\caption{Momentum distribution of the spectral density \eqref{spectral} along the $k_x$ axis (left panel) and the $k_y$ axis (right panel) for varying values of the amplitude of the ionic lattice with the wave vector $p_I=3.6$. 
Translational invariance is broken explicitly in the $x$ direction and the first Brillouin zone boundary is at $k_x=1.8$. 
The red and purple curves correspond to $a_0=0.1$ and $a_0=0.9$, respectively. 
We have fixed $\omega=10^{-6}$, $q=2$, $T=0.0069$. 
Since the vertical axis is logarithmic, the two curves describe very sharp peaks indicative of a Fermi surface.}
\label{fig:fdAvskionicq20k36}
\end{center}
\end{figure}

Recall that this case was described by a spatially varying chemical potential, 
\begin{equation}\label{ioniclatticeapp}
\mu(x)=A_t(z=1,x)=\mu[1+a_0 \cos(p_I\, x)]\,,
\end{equation}
used to represent the potential felt by electrons in an array of ions.
Figs. \ref{fig:fdAvskionicq20}, \ref{fig:fdAvskionicq23} and 
\ref{fig:fdAvskionicq15} and \ref{fig:fdAvskionicq20k36} are the analog of, 
respectively, Figs. \,\ref{fig:Avskionicq20}, \,\ref{fig:Avskionicq23} \,\ref{fig:Avskionicq15} and \ref{fig:Avskionicq20k36} but obtained using  
the reduced zone representation~\eqref{spectral}.


\subsection{Scalar lattice case}

\begin{figure}[ht!]
\begin{center}
\includegraphics[width=.485\textwidth]{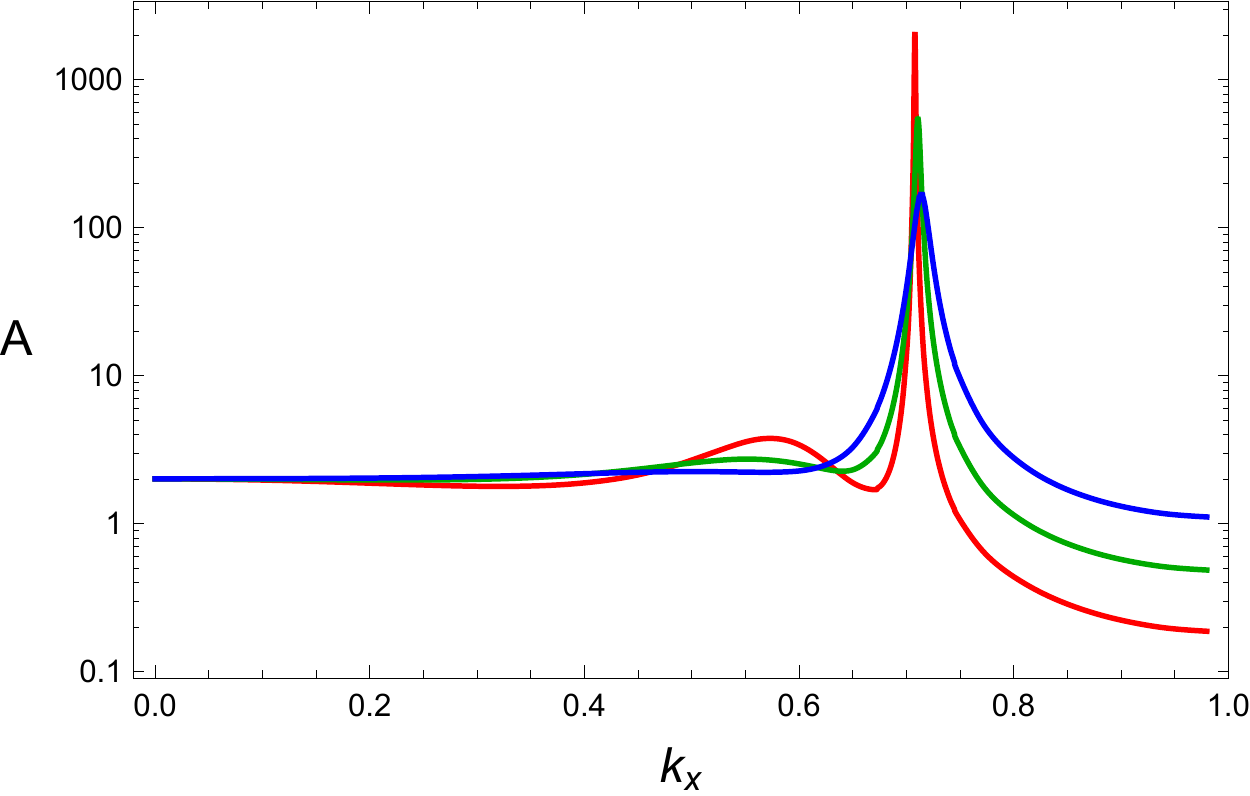}\quad
\includegraphics[width=.485\textwidth]{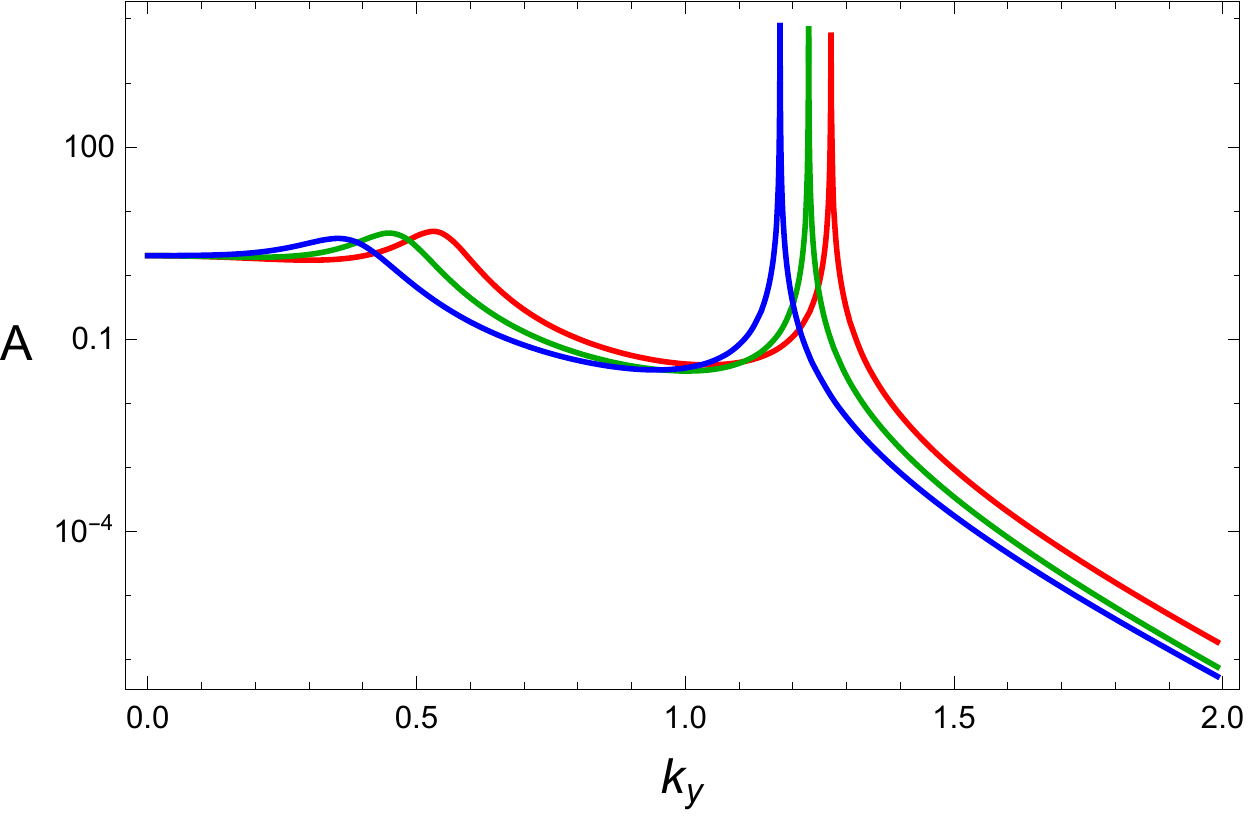}
\caption{Momentum distribution of the spectral density \ref{fig:Avskscalar} along the $k_x$ axis (left panel) and the $k_y$ axis (right panel) for varying values of the amplitude of the scalar lattice. 
Translational invariance is broken explicitly in the $x$ direction and the first Brillouin zone boundary is at $k_x=1$.  
The vertical axis is logarithmic. 
In both plots the red, green and blue curves correspond to $A_0=3, 6$ and $11$, respectively. 
We have chosen $\omega=10^{-6}$, $q=2$, $T=0.0069$ and wave vector $p_S=1$.}
\label{fig:fdAvskscalar}
\end{center}
\end{figure}

The scalar lattice was introduced by adopting 
spatially inhomogeneous boundary condition for the scalar field $\phi$ in the matter sector (\ref{scalarlagmain}), given by
\begin{equation}
\label{scalarlatticeapp}
\phi_s(x)=A_0\cos(p_S\, x)\,,
\end{equation}
and corresponding to a periodic source for the dual scalar operator.
Fig. \ref{fig:fdAvskscalar} represents the analog of Fig. \ref{fig:Avskscalar}, computed using 
the spectral function \eqref{spectral}.

\subsection{Pair Density Wave Case}

\begin{figure}[ht!]
\begin{center}
\includegraphics[width=.49\textwidth]{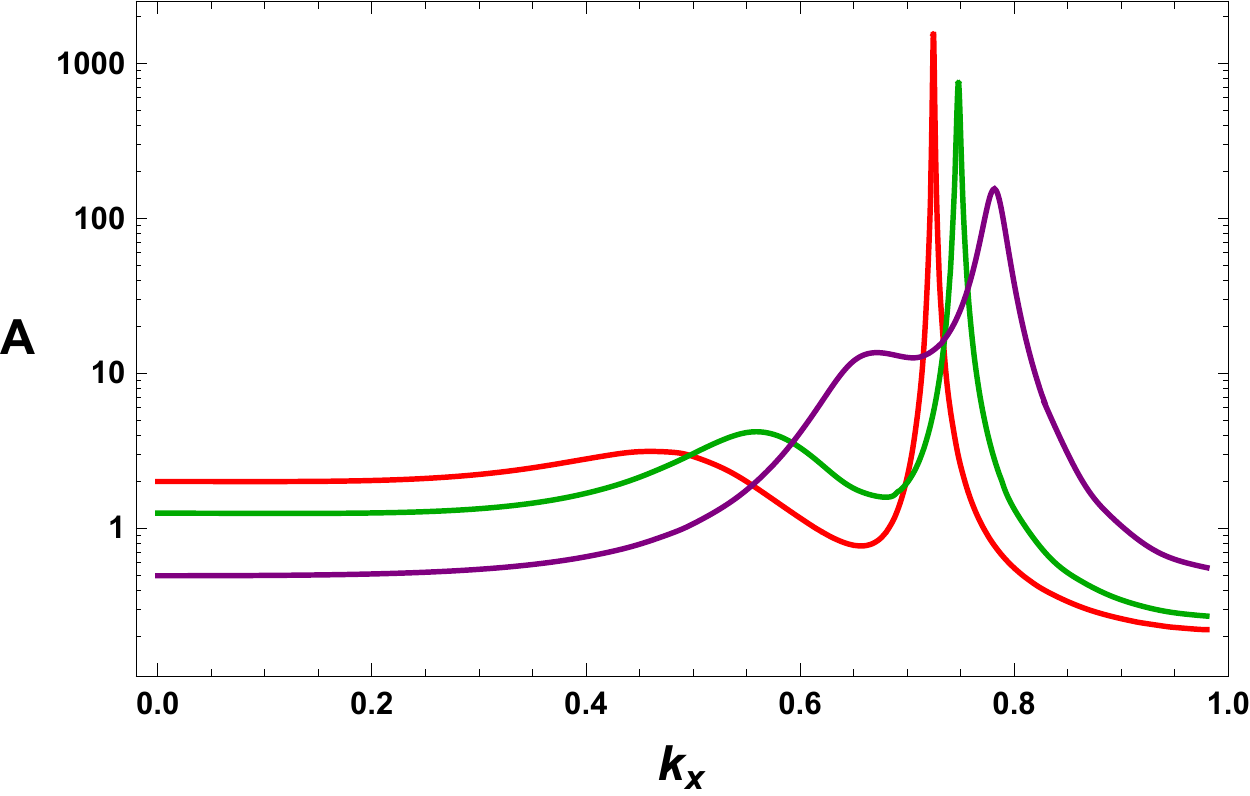}\quad
\includegraphics[width=.48\textwidth]{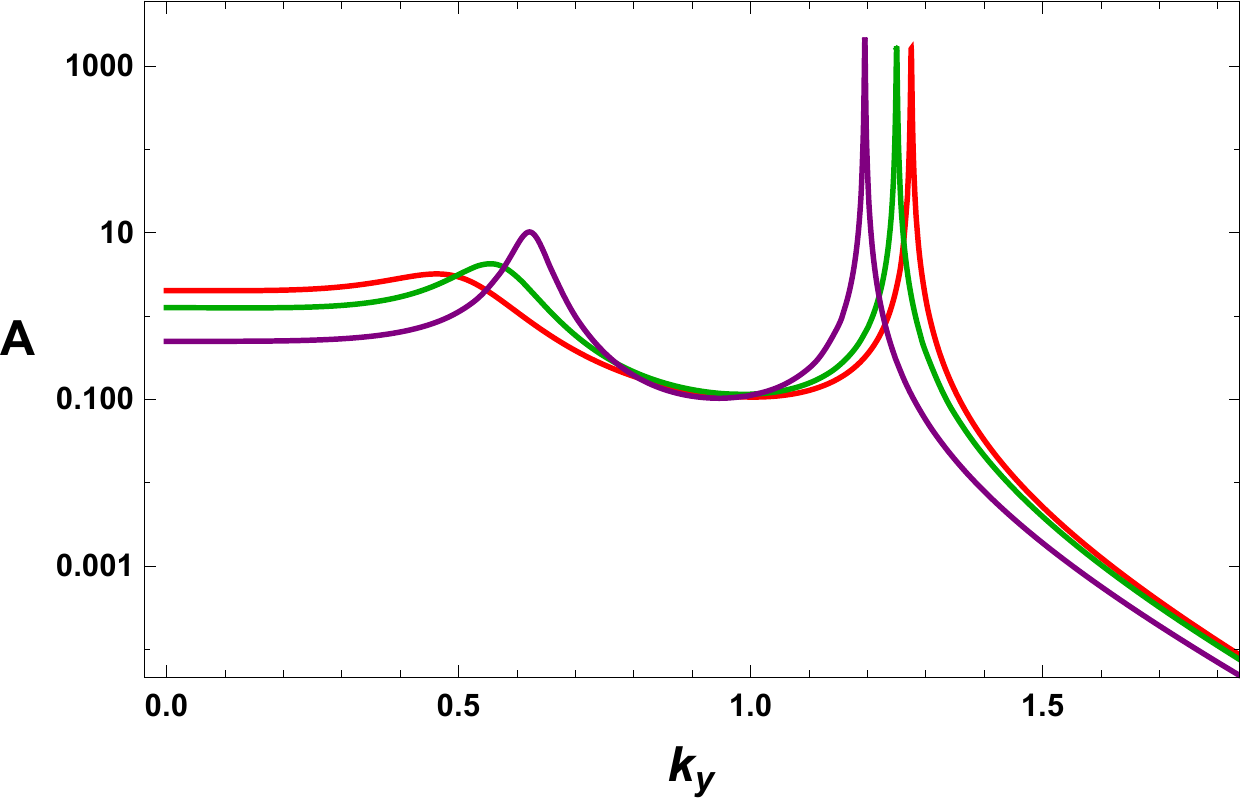}
\caption{Momentum distribution of the spectral density \eqref{spectral} along the $k_x$ axis (left panel) and along the $k_y$ axis (right panel) for varying values of the coupling constant $n$. 
Translational invariance is broken spontaneously along the $x$ direction and the first Brillouin zone boundary is at $k_x=1$.  
In both plots the red curve corresponds to $n=0$, while the green, and purple curves correspond to $n=10$ and $30$, respectively. We have fixed $\omega=10^{-6}$, $q=2$ and chosen $T=0.014$ for the PDW geometry. }
\label{fig:fdAvskPDW}
\end{center}
\end{figure}

In the PDW case, which corresponds to the choice of couplings \ref{PDWcouplings}
for the scalar sector of \eqref{generalmodel}, the U(1) symmetry and translational invariance
are both broken spontaneously.
Fig. \ref{fig:fdAvskPDW} below corresponds to Fig.\,\ref{fig:AvskPDW}, but using the reduced zone representation~\eqref{spectral}.

\subsection{Charge Density Wave Case}

\begin{figure}[H]
\begin{center}
\includegraphics[width=.49\textwidth]{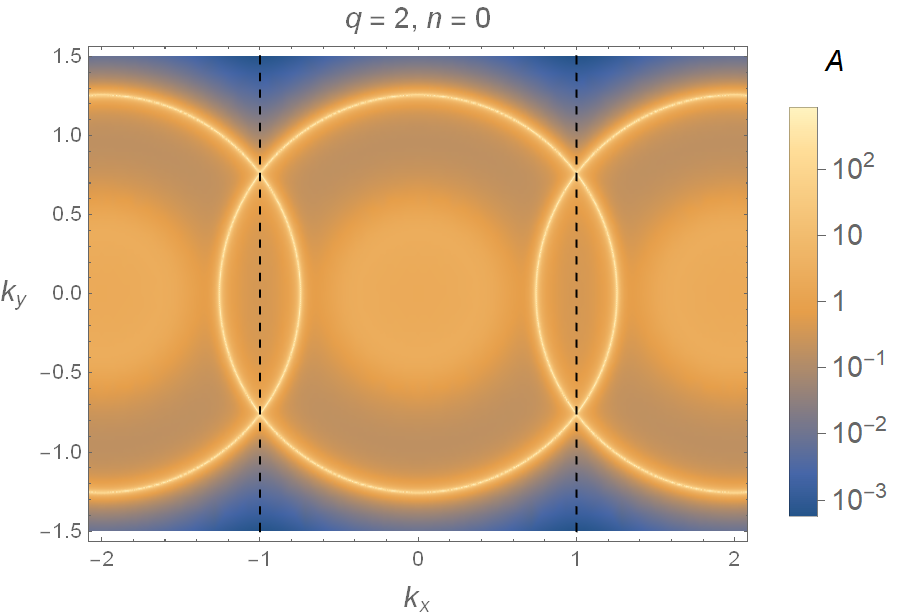}\;\;
\includegraphics[width=.49\textwidth]{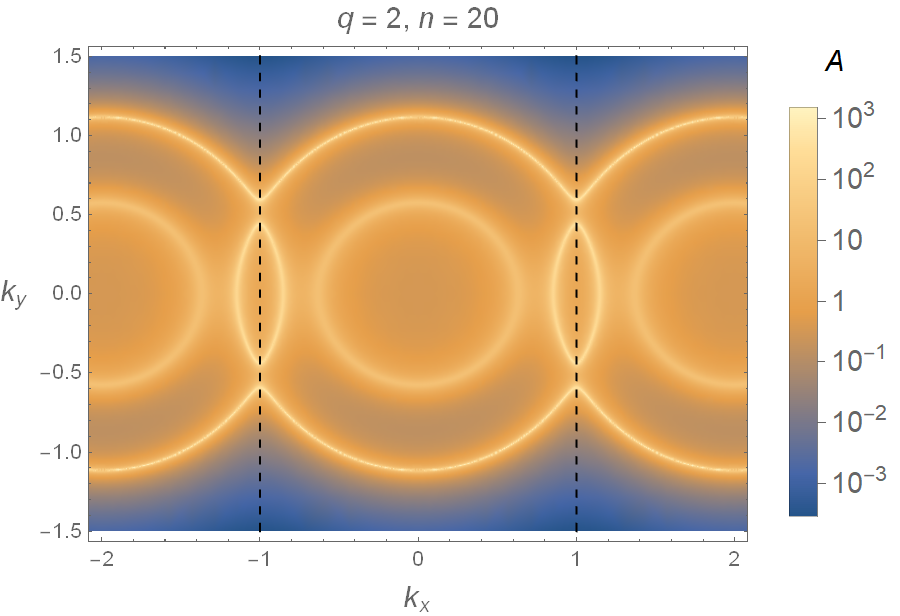}\\
\vspace{10pt}
\includegraphics[width=.49\textwidth]{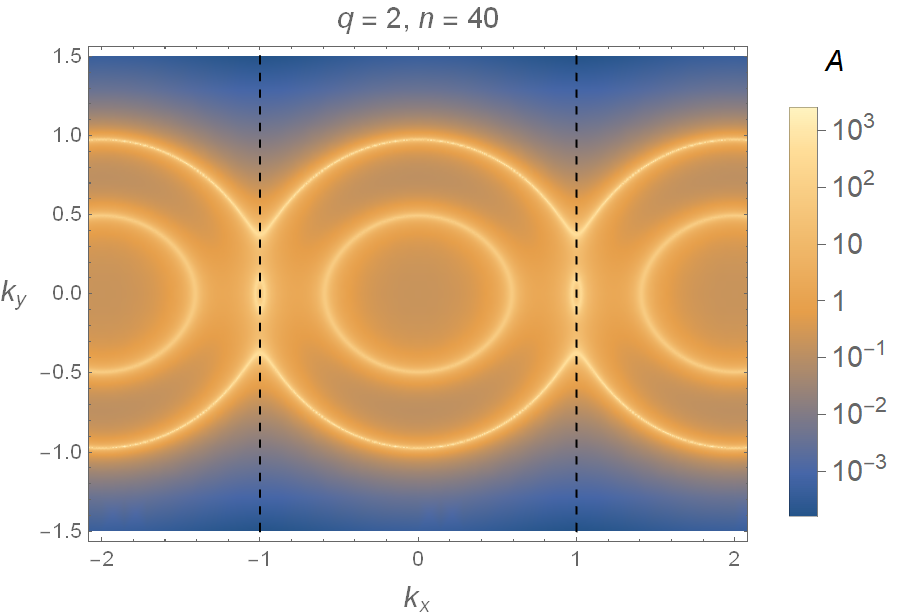}\;\;
\includegraphics[width=.49\textwidth]{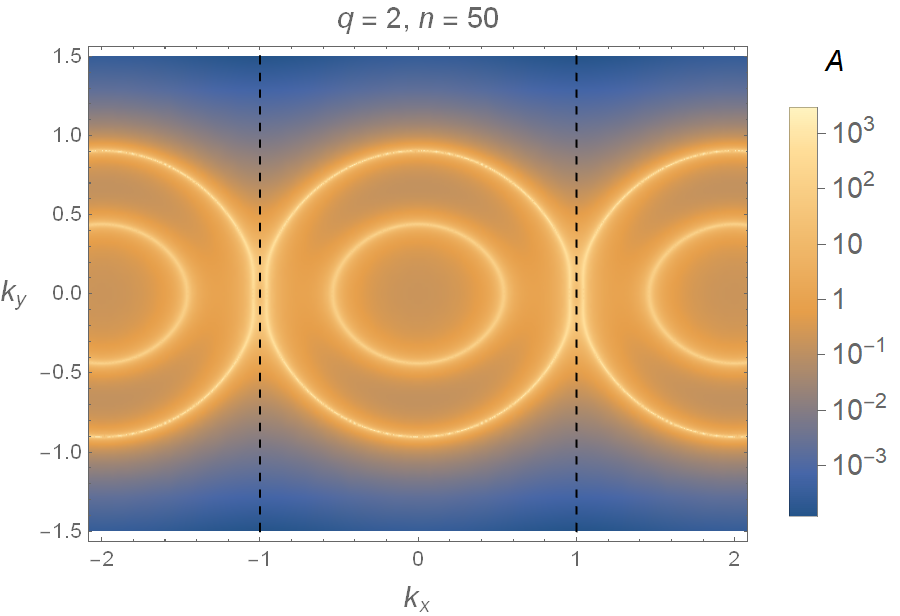}
\caption{Density plot of the momentum distribution of the spectral function \eqref{spectral} in the $(k_x, k_y)$ plane for $q=2$ and $\omega=10^{-6}$. The brightest points denote the location of the Fermi surface. We have periodically extended the data from the first Brillouin zone to the other ones. 
The first Brillouin zone boundary is denoted by the vertical dashed lines at $k_x=\pm 1$, and the CDW geometry has $T=0.02144$ and $k=1$. The four plots correspond to $n=0$, $20$, $40$ and $60$. }
\label{fig:fdFS_CDW_q2}
\end{center}
\end{figure}

\begin{figure}[ht!]
\begin{center}
\includegraphics[width=.49\textwidth]{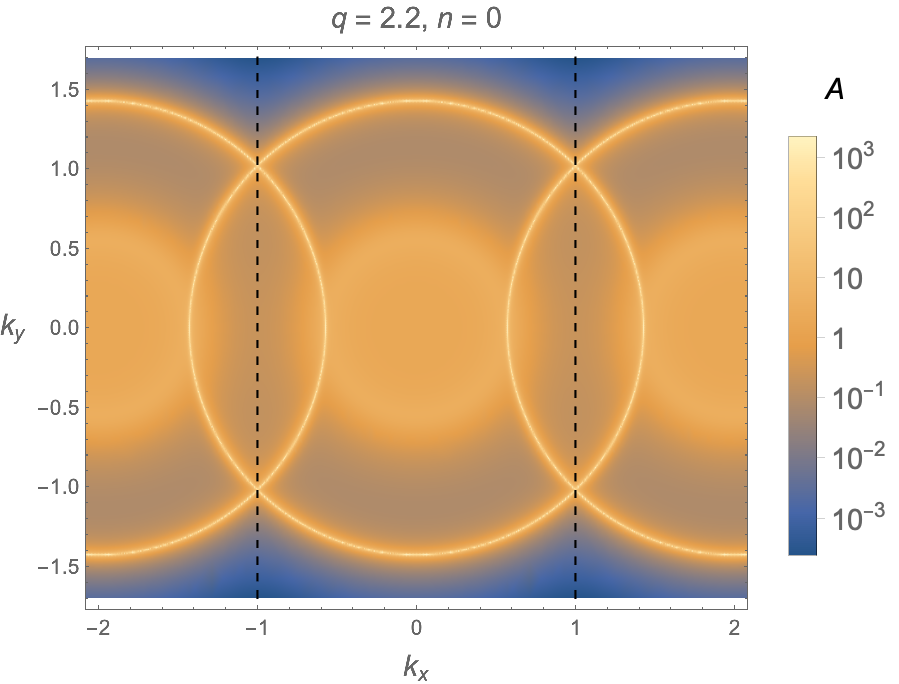}\;\;
\includegraphics[width=.49\textwidth]{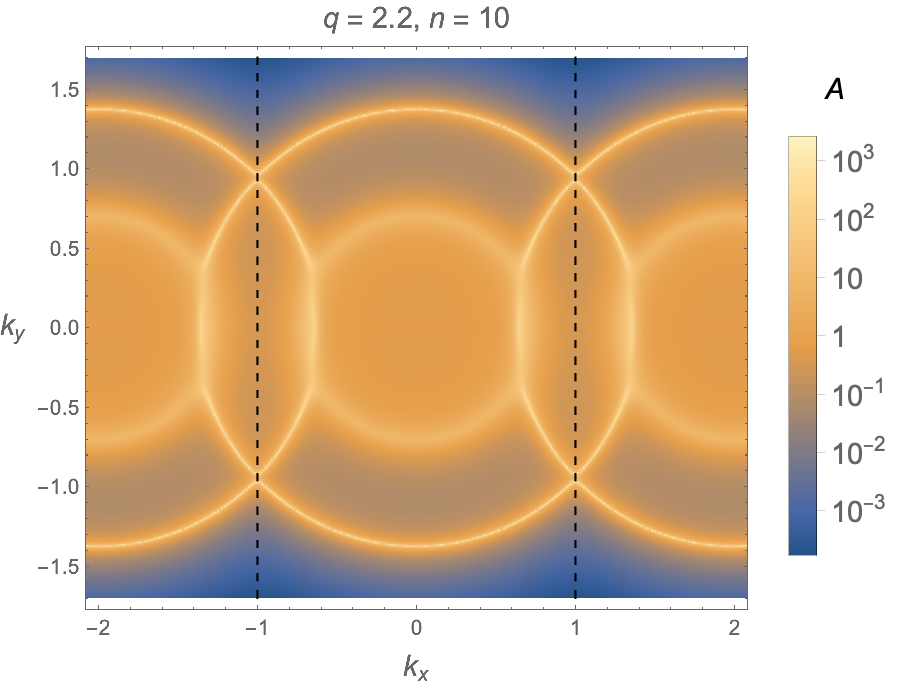}\\
\vspace{10pt}
\includegraphics[width=.49\textwidth]{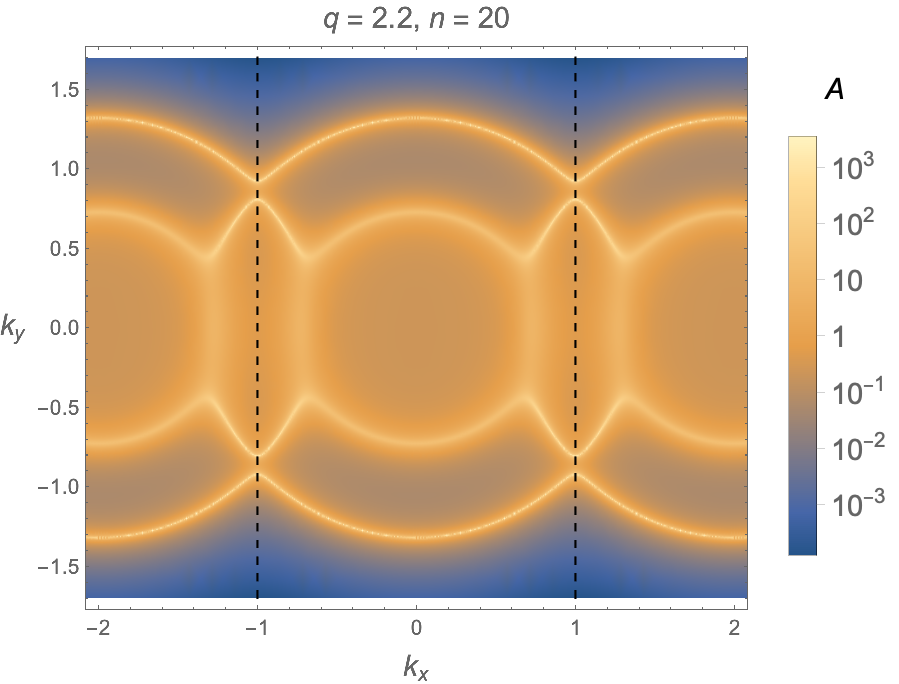}\;\;
\includegraphics[width=.49\textwidth]{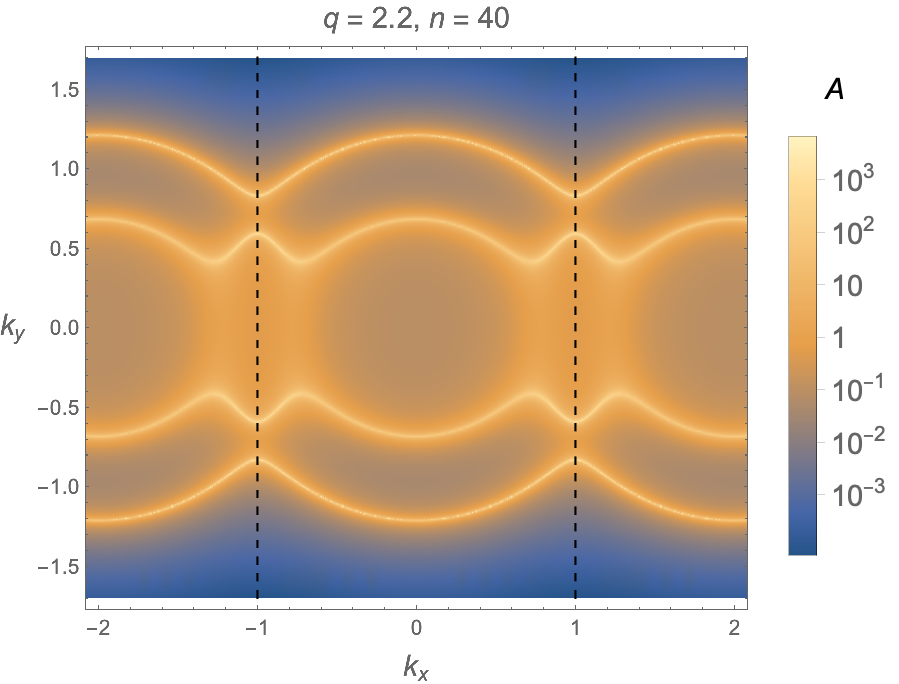}
\caption{Density plot of the momentum distribution of the spectral function~\eqref{spectral}
in the $(k_x, k_y)$ plane for $q=2.2$ and $\omega=10^{-6}$. The brightest points denote the location of the Fermi surface. We have periodically extended the data from the first Brillouin zone to the other ones.
The first Brillouin zone boundary is denoted by the vertical dashed lines at $k_x=\pm 1$, and the CDW geometry has $T=0.02144$ and $k=1$. The four plots correspond to $n=0$, $10$, $20$ and $40$. 
As $n$ increases, some of the vertical parts of the Fermi surface gradually vanish.}
\label{fig:fdFS_CDW_q22}
\end{center}
\end{figure}

\begin{figure}[ht!]
\begin{center}
\includegraphics[width=.49\textwidth]{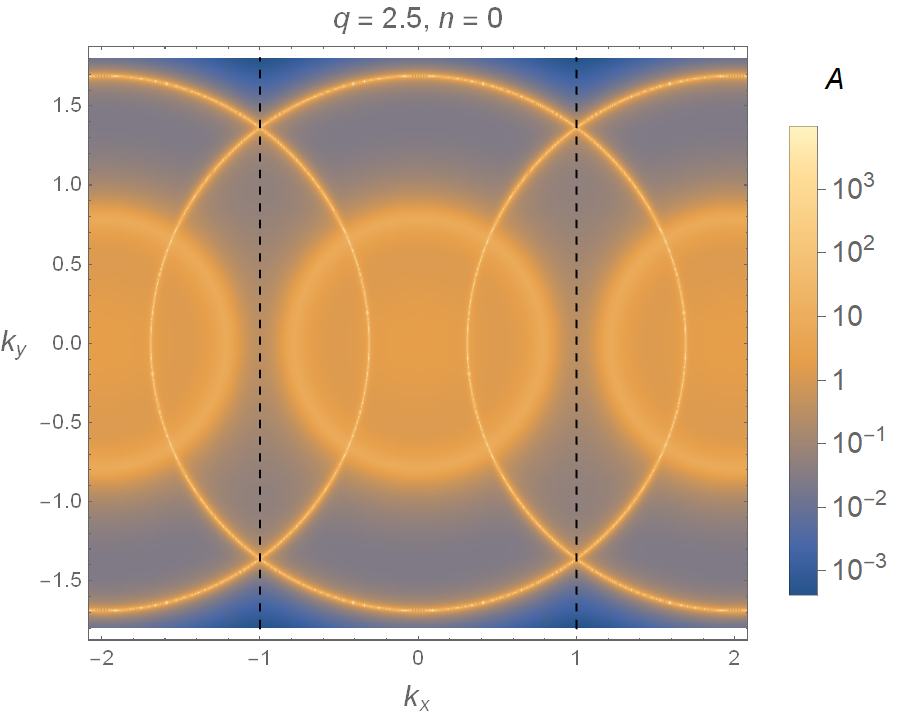}\;\;
\includegraphics[width=.49\textwidth]{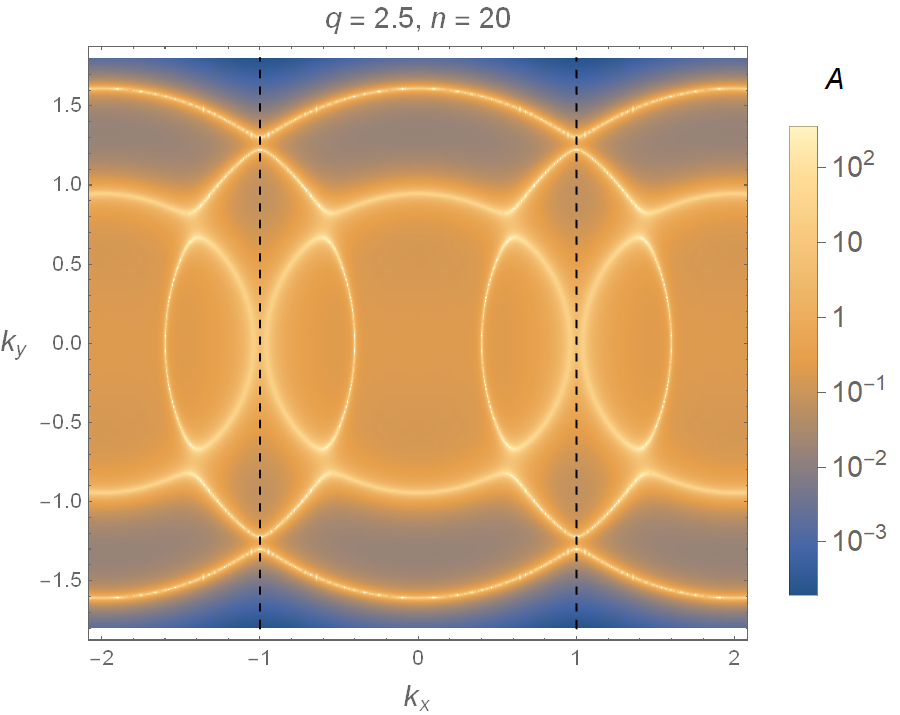}\\
\vspace{10pt}
\includegraphics[width=.49\textwidth]{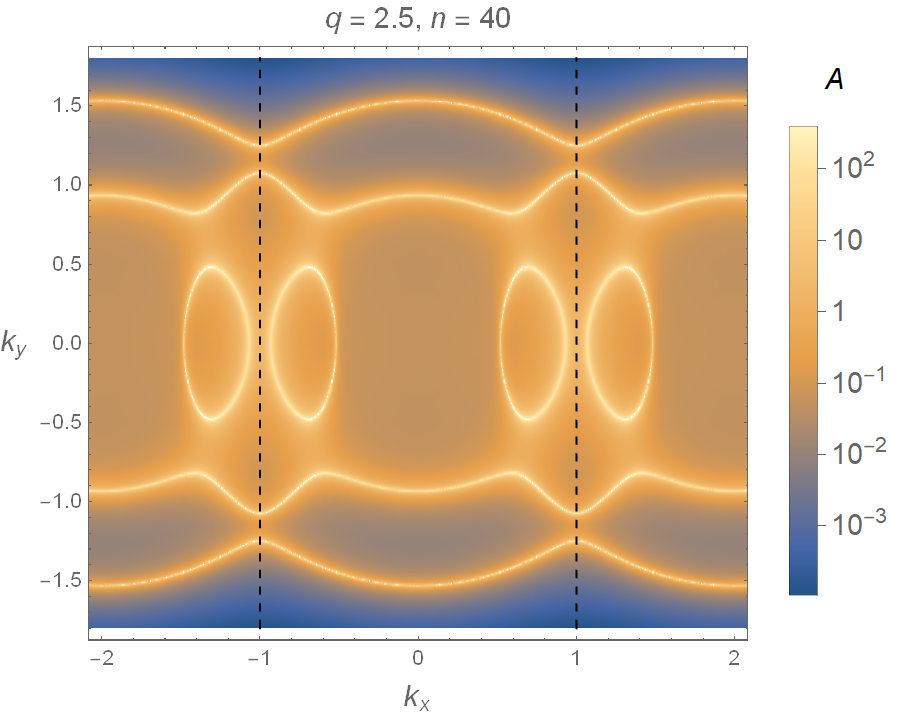}\;\;
\includegraphics[width=.49\textwidth]{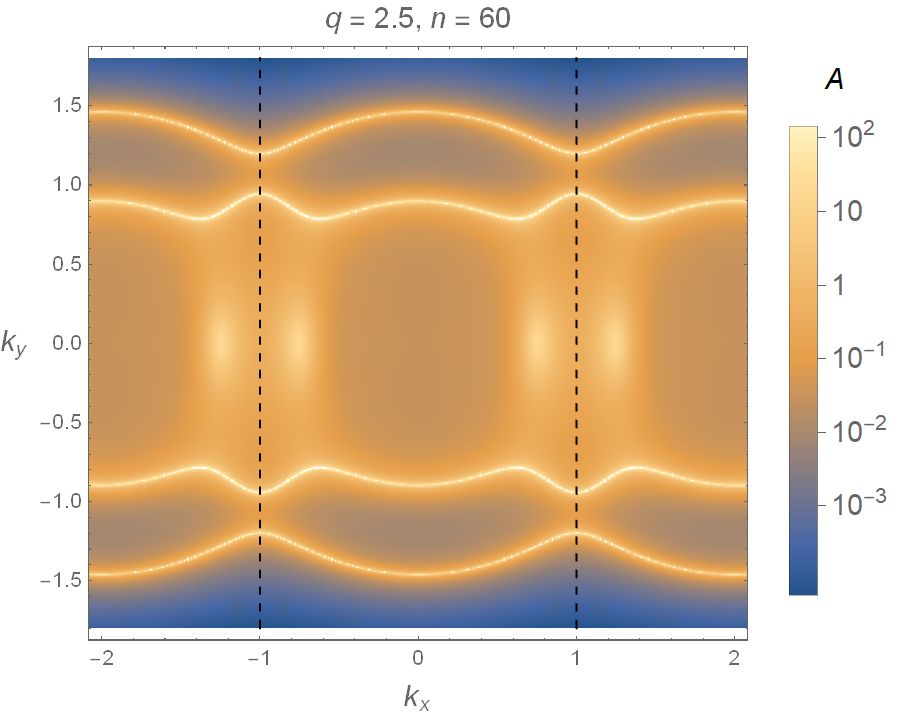}
\caption{Density plot of the momentum distribution of the spectral function \eqref{spectral} in the $(k_x, k_y)$ plane for $q=2.5$ and $\omega=10^{-6}$. The brightest points denote the location of the Fermi surface. We have periodically extended the data from the first Brillouin zone to the other ones.
The first Brillouin zone boundary is denoted by the vertical dashed lines at $k_x=\pm 1$, and the CDW geometry has $T=0.02144$ and $k=1$. The four plots correspond to $n=0$, $20$, $40$, and $60$. 
As $n$ increases, some of the vertical parts of the Fermi surface gradually vanish.}
\label{fig:fdFS_CDW_q25}
\end{center}
\end{figure}

\begin{figure}[ht!]
\begin{center}
\includegraphics[width=.48\textwidth]{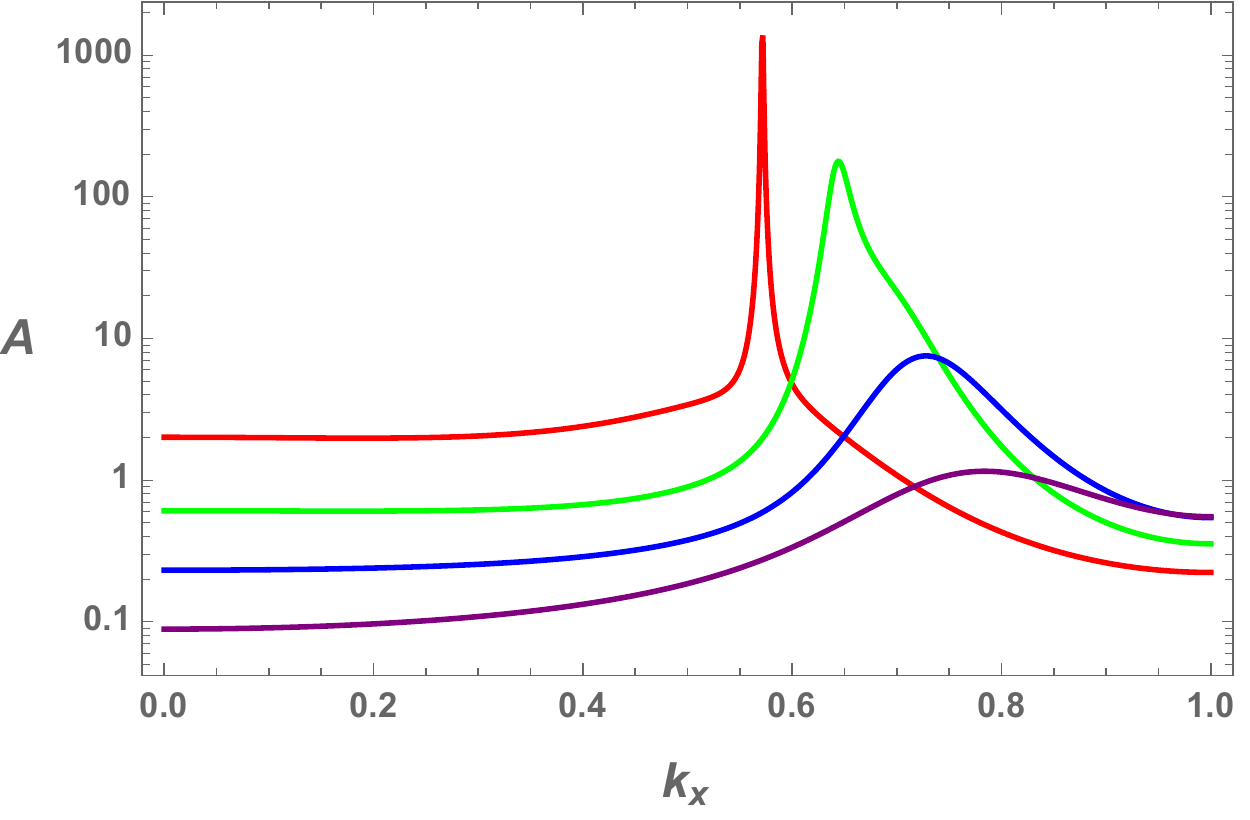}\quad
\includegraphics[width=.48\textwidth]{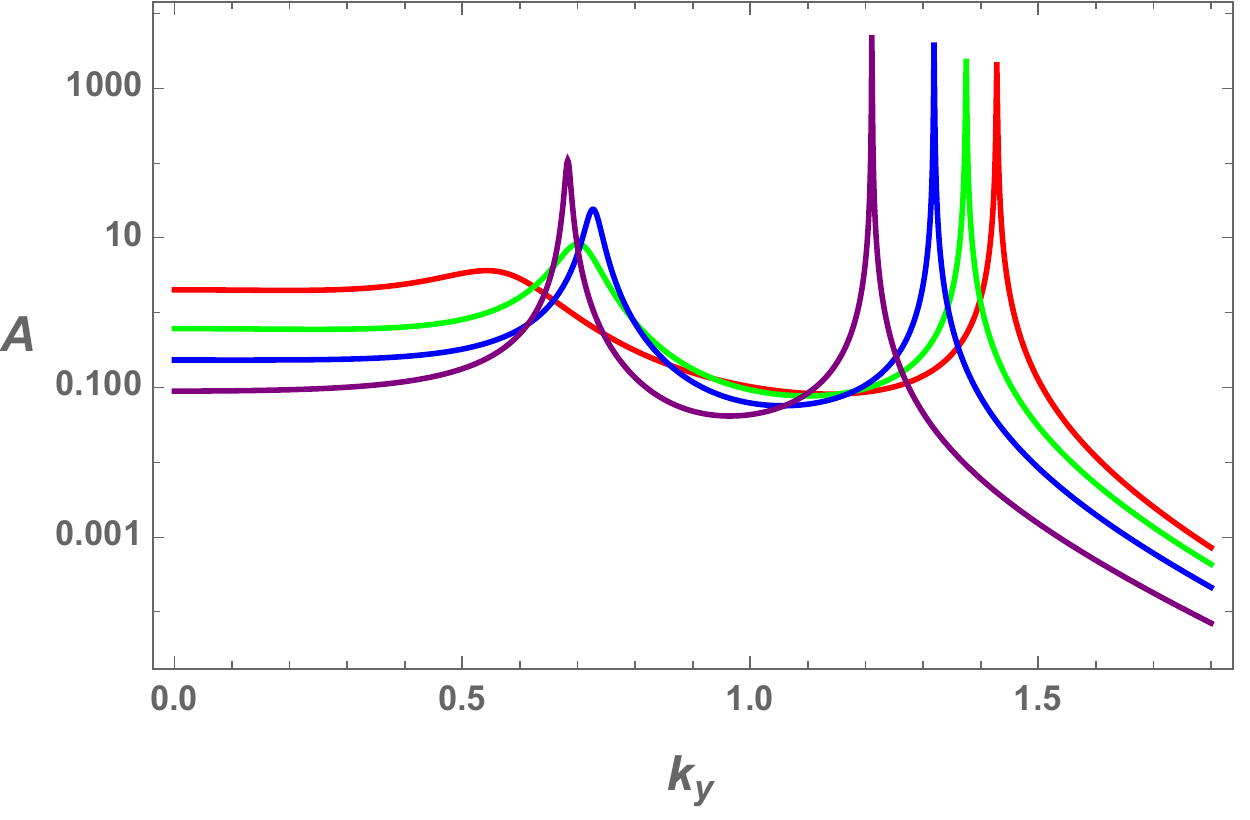}
\caption{Momentum distribution of the spectral density~\eqref{spectral} along the $k_x$ axis (left panel) for $k_y=0$ and along the $k_y$ axis (right panel) for $k_x=0$ for varying values of the coupling constant $n$. 
Translational invariance is broken spontaneously along the $x$ direction and the first Brillouin zone boundary is at $k_x=1$. In both plots the red curve corresponds to $n=0$, while the green, blue and purple curves correspond to $n=10, 20$ and $40$, respectively. We have fixed $\omega=10^{-6}$, $q=2.2$ and chosen $T=0.02144$ for the CDW geometry. }
\label{fig:fdAvskCDW}
\end{center}
\end{figure}

We conclude with the CDW case, in which the U(1) symmetry is preserved ($q_A=q_B=0$ and $\theta$ truncated out).
Figs.~\ref{fig:fdFS_CDW_q2}, \ref{fig:fdFS_CDW_q22} and~\ref{fig:fdFS_CDW_q25}  
are density plots of the momentum distribution of the spectral function~\eqref{spectral}
corresponding to $q=2$, $q=2.2$ and $q=2.5$, respectively (the first two plots of Fig.\,\ref{fig:fdFS_CDW_q2} can be compared to Fig.\,5 and Fig.\,9 of our previous paper~\cite{Cremonini:2018xgj})). They are the analogs of Figs.\,\ref{fig:FS_CDW_q2}, \ref{fig:FS_CDW_q22} 
and \ref{fig:FS_CDW_q25}, but using the reduced zone representation.
Similarly, Fig. \ref{fig:fdAvskCDW} shows the momentum distribution of the spectral density (along the $k_x$ axis in the left panel and along the $k_y$ axis in the right panel)  for varying values of the coupling constant $n$ and is the analog of 
Fig.\,\ref{fig:AvskCDW}.
%

%
%
%
%
%

%
%

\section{Energy Distribution of the Spectral Density}
\label{app:energy}

\begin{figure}[ht!]
\begin{center}
\includegraphics[width=.49\textwidth]{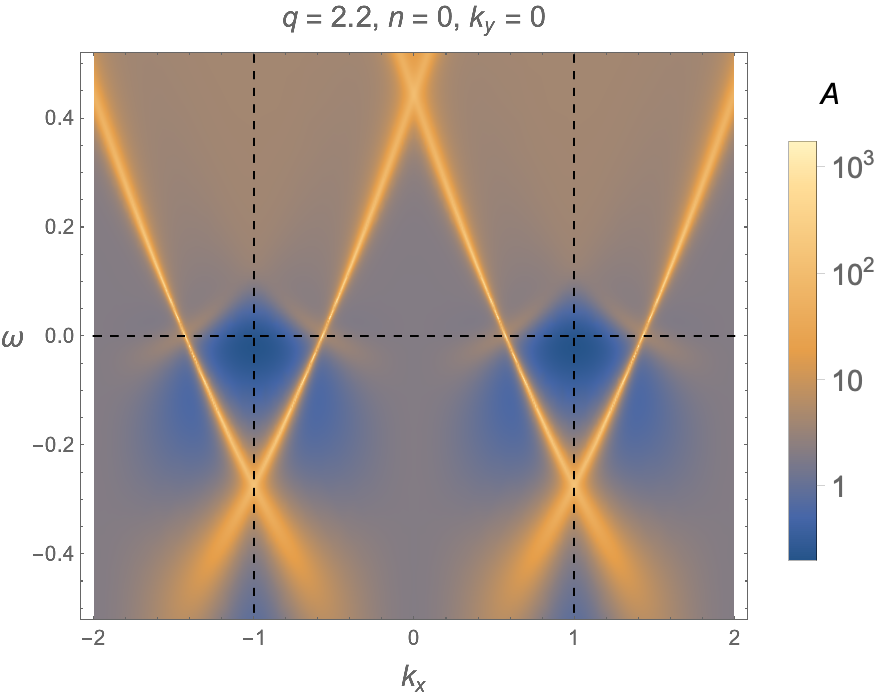}
\includegraphics[width=.49\textwidth]{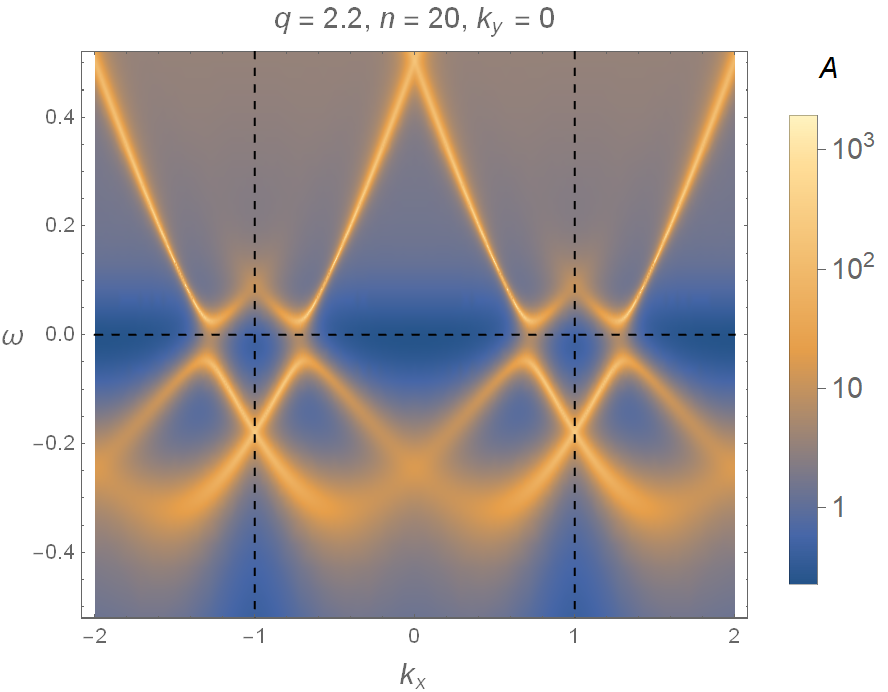}
\caption{Density plots of the energy distribution of the spectral function~\eqref{spectral} along the $k_x$ axis. The intersections between the horizontal dashed line ($\omega=0$) and the brightest points correspond to the location of the Fermi surface. In the right plot there is an ``energy gap" opening near $\omega=0$ due to increasing the lattice strength. We have periodically extended the data from the first Brillouin zone to the other ones.
The first Brillouin zone boundary is denoted by the vertical dashed lines at $k_x=\pm 1$, and the CDW background geometry has $T=0.02144$ and $k=1$.}
\label{fig:DISP_CDW_q22}
\end{center}
\end{figure}

In this section we include preliminary results on the energy distribution function, i.e. the spectral density as a function of $\omega$, to shed some light on the origin of the Fermi surface segmentation. 
In appendix A of our previous paper~\cite{Cremonini:2018xgj} we showed the energy distribution function for different choices of momentum.
We were interested in how the spectral density evolves as the momentum is varied, in order to probe the formation and evolution of the Fermi surface itself.
One interesting feature we observed~\cite{Cremonini:2018xgj}, which appears to be in agreement with ARPES experiments on the cuprates, is the ``peak-dip-hump" structure in the energy distribution function at fixed momentum.

Here we focus on the CDW case and adopt the reduced zone representation~\eqref{spectral}. 
Although  there is some tension between the two representations,  as we have discussed in previous sections, 
it turns out to be more convenient to use the reduced zone representation in order to see the structure of the peaks in the spectral function.
The left plot of Fig.\,\ref{fig:DISP_CDW_q22} shows the energy distribution of the spectral density for $q=2.2$, $n=0$ (corresponding to the first plot of Fig.\,\ref{fig:FS_CDW_q22} as well as Fig.\,\ref{fig:fdFS_CDW_q22}). 
The intersections between the horizontal dashed line (at $\omega=0$) and the brightest lines give the location of the Fermi surface. 
One finds that near the Fermi surface the dispersion relation along the $k_x$ direction behaves quite linearly. 

The energy distribution for $n=20$ is shown in the right plot of Fig.\,\ref{fig:DISP_CDW_q22}, which corresponds to the third plot of Fig.\,\ref{fig:FS_CDW_q22} as well as Fig.\,\ref{fig:fdFS_CDW_q22}. 
We now see the opening an ``energy gap" near $\omega=0$, due to the increase in the strength of the lattice. 
Indeed, in contrast to the $n=0$ case, there is no intersection between the $\omega=0$ horizontal dashed line and the brightest points. Thus, there is no longer a Fermi surface at that particular 
location. 
We note that the shape of the dispersion relation is now quite non-trivial, and is complicated by the presence of multiple bands. There is also a sharp asymmetry between positive and negative frequency modes.
As the lattice strength (controlled by $n$) 
is increased, two different effects seem to be at play. One is the lifting of a single band above the Fermi level, the other is the formation of additional bands. The interplay between these two effects then leads to the opening of the band gap, which in turn is associated with the disappearance of a particular segment of the Fermi surface.
We postpone a more detailed analysis of this behavior to future work.

\end{document}